\documentclass{article}

\usepackage{cite}
\usepackage{epsfig}

\def\tr{{\rm tr\,}}
\def\det{{\rm det\,}}

\def\bea{\begin{eqnarray}}
\def\eea{\end{eqnarray}}

\def\dash{\,\textendash\, }
\def\lmatrix{\left(\begin{array}}
\def\rmatrix{\end{array}\right)}

\begin{document}

\vspace{-2cm}

\hfill UCSD/PTH 09-07

\vspace{1cm}

\begin{center}{\Large\bf Chiral properties of $SU(3)$ sextet fermions }\\
\vspace{1cm}
Zolt\'an Fodor$^{\,abcd}$,$\;$ Kieran Holland$^{\,e}$,$\;$ Julius Kuti$^{\,f}$,\\

\vspace{0.2cm}

D\'aniel N\'ogr\'adi$^{\,f}\,$ and$\,$ Chris Schroeder$^{\,f}$

\vspace{0.7cm}

\end{center}

\hspace{3.0cm} $^a${\em Department of Theoretical Physics, University of Wuppertal}

\hspace{3.1cm} {\em Gau$\beta$strasse 20, Wuppertal 42119, Germany}

\vspace{0.2cm}

\hspace{3.0cm} $^b${\em NIC/DESY Zeuthen Forschungsgruppe}

\hspace{3.1cm} {\em Platanenalle 6, Zeuthen 15738, Germany }

\vspace{0.2cm}

\hspace{3.0cm} $^c${\em Forschungszentrum J\"ulich, D-52425 J\"ulich, Germany }

\vspace{0.2cm}

\hspace{3.0cm} $^d${\em Institute for Theoretical Physics, E\"otv\"os University}

\hspace{3.1cm} {\em P\'azm\'any P\'eter s\'et\'any 1/A, Budapest 1117, Hungary }

\vspace{0.2cm}

\hspace{3.0cm} $^e${\em Department of Physics, University of the Pacific}

\hspace{3.2cm}{\em 6301 Pacific Avenue, Stockton, CA 95211, USA}

\vspace{0.2cm}

\hspace{3.0cm} $^f${\em Department of Physics, University of California San Diego}

\hspace{3.1cm} {\em 9500 Gilman Dr, La Jolla, CA 92093, USA }

\vspace{0.5cm}

\begin{center}
Emails

\vspace{0.1cm}

fodor@bodri.elte.hu, kholland@pacific.edu, jkuti@ucsd.edu \\ nogradi@bodri.elte.hu, crs@physics.ucsd.edu
\end{center}

\vspace{1cm}

\begin{abstract}
$SU(3)$ gauge theory with overlap fermions in the 2-index symmetric (sextet) and fundamental representations is considered.
A priori it is not known what the pattern of chiral symmetry breaking is in a higher dimensional representation although the
general expectation is that if two representations are both complex, the breaking pattern will be the same. This expectation
is verified for the sextet at $N_f = 0$ in several exact zero mode sectors. It is shown that if the volume is large enough 
the same random matrix ensemble describes both the sextet and fundamental Dirac eigenvalues. The number of zero modes for the
sextet increases approximately 5-fold relative to the fundamental in accordance with the index theorem for small lattice spacing 
but zero modes which do not correspond to integer topological charge
do exist at larger lattice spacings. The zero mode number dependence of the random matrix model predictions correctly match
the simulations in each sector and each representation.
\end{abstract}

\newpage

\section{Introduction}

An elegant and attractive way of solving the hierarchy problem of the Standard Model is the technicolor paradigm
\cite{Weinberg:1975gm, Susskind:1978ms}, for a review and additional references see \cite{Hill:2002ap}. The basic building block of any
technicolor model is a non-abelian gauge theory coupled to some number of fermion flavors transforming in some representation
of the gauge group. The renormalized coupling towards the infrared may run similarly to QCD, may exhibit a walking behavior or
may run into a fixed point. Generically, it is possible to realize all three possibilities for a given gauge group by
changing the representation and/or flavor number. If both the gauge group and the representation are fixed the flavor number will
determine in which phase the gauge theory is in the infrared.

Chiral symmetry breaking is intimately tied to the infrared behavior. Clearly, if chiral symmetry is broken, the theory can
not have a non-trivial conformal fixed point. At $N_f = 0$ chiral symmetry is broken and as $N_f$ is increased at a critical $N_f$
the theory will enter the conformal window. It is generally believed that the restoration of chiral symmetry will signal 
the lower edge of the conformal window. Strictly
speaking this is not necessarily the case because in principle chiral symmetry can be intact while the theory is not conformal (for example
by generating a dynamical scale not in the chiral condensate but another quantity) but the general expectation is that this 
exotic scenario will not take place.

In this work the 2-index symmetric (sextet) representation of $SU(3)$ is considered in the quenched approximation. This model with $N_f = 2$ fermions is
attractive for phenomenology for several reasons. First, since the sextet is a complex representation, if chiral symmetry breaking
does take place it is expected on general grounds to give rise to 3 Goldstone modes which will be eaten appropriately by the $W$ and $Z$ bosons. 
Second, the $N_f = 2$
value might not be far from the lower edge of the conformal window and the coupling might exhibit the necessary walking behavior.
Third, the perturbative $S$-parameter is rather small making it plausible that the model can be made consistent with electro-weak
precision data. The likely smallness of the $S$-parameter is a consequence of the fact that the fermion content is increased by
increasing the dimension of the representation and not the flavor number \cite{Sannino:2004qp, Dietrich:2006cm}.
All of the above, of course, is only possible if the theory is below the conformal window. If chiral symmetry
breaking does take place, this is clearly the case. This is our primary motivation for the current study of chiral properties of
sextet fermions. 

Overlap fermions \cite{Neuberger:1997fp} preserve chiral symmetry exactly and possess exact zero modes at finite lattice spacing. The zero modes are
indicative of the topology of the underlying gauge field while near-zero modes give rise to chiral symmetry breaking via the
Banks-Casher relation \cite{Banks:1979yr}. In the $\varepsilon$-regime random matrix theory (RMT) \cite{Shuryak:1992pi,
Verbaarschot:1993pm, Verbaarschot:1994qf, Halasz:1995qb} gives detailed predictions for the statistical
properties of non-zero modes. How well RMT actually describes the non-zero modes of the sextet overlap operator will be
addressed in this work at $N_f = 0$. Our findings for the zero-modes and the associated questions about topology have been
reported in \cite{Fodor:2009nh}.

A priori it is not known what the pattern of chiral symmetry breaking is in an arbitrary representation. On general grounds one
expects the sextet breaking to be the same as for the fundamental representation because both representations are complex. This expectation is
verified here by the observation that the same RMT ensemble, the chiral unitary one, describes both the fundamental and the sextet
eigenvalues in each exact zero mode sector $\nu$. The number of zero modes depends on the representation and once a gauge
configuration is given the number of sextet zero modes is approximately $5$ times the number of fundamental zero modes as dictated by the
index theorem. The deviation from this rule is a lattice artifact which becomes less and less for decreasing lattice spacing
\cite{Fodor:2009nh}. The RMT predictions are sensitive to $\nu$ and the matching of the simulation to RMT with the correct number
of $\nu$ is a highly non-trivial test. A staggered fermion study necessarily limited to the trivial sector has been reported in
\cite{Damgaard:2001fg} and an overlap study with two different $SU(2)$ representations and the fundamental of $SU(3)$ appeared in
\cite{ }.

Previous lattice studies with unusual fermion content include gauge groups $SU(2)$ and $SU(3)$ and various representations
\cite{Kogut:1982fn, Kogut:1984sb, Iwasaki:1991mr,
Damgaard:1997ut, Karsch:1998qj, Damgaard:2001fg, Iwasaki:2003de, Catterall:2007yx, Appelquist:2007hu, DeGrand:2008kx, Shamir:2008pb, DelDebbio:2008tv,  
DelDebbio:2008zf, DelDebbio:2009fd, Fodor:2008hn, Fodor:2008hm, Catterall:2008qk,  Hietanen:2008mr, Deuzeman:2008sc,
Armoni:2008nq, Jin:2008rc, Appelquist:2009ty, 
Hietanen:2009az, Hietanen:2009zz, Deuzeman:2009mh, Fodor:2009nh, DeGrand:2009et, Hasenfratz:2009ea, Fodor:2009wk, Nagai:2009ip} which were largely motivated
by the same considerations as mentioned here, with some exceptions.

\section{Dirac spectrum}

The low lying Dirac spectrum is indicative of the infrared properties of the theory. It will exhibit markedly
different behavior if chiral symmetry is spontaneously broken or if the infrared dynamics is conformal. Using the Dirac spectrum
as a tool to decide in which phase a gauge theory is has been suggested first in \cite{Fodor:2008hm} and later this idea was 
applied in \cite{DeGrand:2009et}.

\subsection{Chiral symmetry breaking and RMT}

The Banks-Casher relation in a theory with spontaneous symmetry breaking relates the low end of the spectral Dirac density to the
chiral condensate,
\bea
\Sigma = \lim_{m\to0} \lim_{V\to\infty} \frac{\pi \rho(0)}{V}\;.
\eea
It also implies that the low-lying eigenvalues are dense in the sense that the average spacing is inversely proportional to the
volume,
\bea
\label{chideltal}
\langle \Delta \lambda \rangle = \frac{\pi}{\Sigma V}\;.
\eea

The low energy dynamics in a finite box of spatial extent $L$ is accurately described by chiral
perturbation theory if $f_\pi L \gg 1$. If in addition the quark mass is chosen such that $m_\pi L \ll 1$ the derivative terms in
the chiral Lagrangian can be dropped. This $\varepsilon$-regime is then analytically calculable in the microscopic limit with the help of RMT which can
also be directly derived from the chiral Lagrangian \cite{Osborn:1998qb}. The comparison is valid in the microscopic large $L$ limit, meaning that the
Dirac eigenvalues $\lambda$ and fermion masses $m$ are rescaled, $\zeta = \lambda \Sigma V$ and $\mu = m \Sigma V$, where $\zeta$
and $\mu$ are kept finite. Hence the comparison is relevant for the small eigenvalues which are $\lambda \sim 1/V$.

The RMT description only depends on the rescaled fermion masses, $N_f$, the number of zero modes $\nu$ and the pattern of
symmetry breaking. Since both the fundamental and sextet representations are complex, the symmetry breaking pattern is expected to
be the same \cite{Dimopoulos:1979sp}.
Then the relevant random matrix model is governed by the following partition function,
\bea
\label{rmtz}
Z = \int dW dW^\dagger \det( D + m )^{N_f} \exp( - N \tr W W^\dagger )\,,
\eea
where the complex matrix $W$ is $N \times (N + \nu)$ and the Dirac operator is,
\bea
D = \lmatrix{cc}  0 & iW \\ iW^\dagger & 0 \rmatrix\,,
\eea
assuming degenerate fermion masses $m$. In the microscopic large $N$ limit the mass $m$ and eigenvalues $\lambda$ are rescaled by $N$,
$\zeta = 2 N \lambda$, $\mu = 2 N m$, similarly to the gauge theory case and then $N \to \infty$. In this limit
the joint distribution of the $k$ smallest eigenvalues,  $\zeta_1 \leq \ldots \leq \zeta_k$, is given by
\cite{Nagao:2000qn, Damgaard:2000ah, Damgaard:2000qt}
\bea
\label{omega}
&& \omega(\zeta_1,\ldots,\zeta_k) = \frac{1}{2} \zeta_k \exp(-\zeta_k^2/4) \prod_{i=1}^{k-1} \left( \zeta_i^{2\nu+1} ( \zeta_i^2+\mu^2)^{N_f} \right)
\prod_{i>j}^{k-1} (\zeta_i^2 - \zeta_j^2 )^2 \mu^{\nu N_f} \times \\
&& \hspace{-1cm} \frac{ {\cal Z}_2( \sqrt{ \mu^2 + \zeta_k^2 },\ldots, \sqrt{ \mu^2 + \zeta_k^2 }, 
\sqrt{\zeta_k^2-\zeta_1^2}, \sqrt{\zeta_k^2-\zeta_1^2}, \ldots \sqrt{\zeta_k^2-\zeta_{k-1}^2}, \sqrt{\zeta_k^2-\zeta_{k-1}^2},
\zeta_k, \ldots, \zeta_k ) }{ {\cal Z}_\nu( \mu,\ldots, \mu ) }
\eea
where $\sqrt{\mu^2 + \zeta_k^2}$ and $\zeta_k$ in the argument of ${\cal Z}_2$ appear $N_f$ and $\nu$ times, respectively,
while $\mu$ appears $N_f$ times in the argument of ${\cal Z}_\nu$. The expression ${\cal Z}_a$ with any number of arguments is given by the ratio,
\bea
{\cal Z}_a(x_1,\ldots,x_m) = \frac{ \det_{1\leq i,j \leq m}\left( x_j^{i-1} I_{a+i-1}(x_j) \right) }{ \prod_{i > j}^m (x_i^2 - x_j^2) }\,,
\eea
which has a finite limit if some of the $x_i$ variables become degenerate. The distribution of the $k^{th}$ eigenvalue is then,
\bea
\label{pzetak}
p_k(\zeta_k) = \int_0^{\zeta_k} d\zeta_1 \int_{\zeta_1}^{\zeta_k} d\zeta_2 \ldots \int_{\zeta_{k-2}}^{\zeta_k} d\zeta_{k-1}
\,\omega(\zeta_1,\ldots,\zeta_k)\,.
\eea

In order to match the gauge theory calculation onto the RMT prediction, the scale $\Sigma$ needs to be determined so that the
gauge theory eigenvalues can be rescaled appropriately by $\Sigma V$. In order to have a completely parameter free prediction we
will consider the distribution of ratios of eigenvalues. The unknown scale $\Sigma$ drops out from these at $N_f = 0$. For example the
distribution of the ratio of the first and second eigenvalue, $r=\lambda_1/\lambda_2=\zeta_1/\zeta_2 < 1$, is simply
\bea
\label{p12}
p_{1,2}(r) = \int_0^\infty d\zeta_2 \, \omega( r \zeta_2, \zeta_2 )\,,
\eea
and can be directly compared with the corresponding distribution in the gauge theory. Similar integrals apply to any ratio $r =
\lambda_j / \lambda_k$.

The universality of the RMT description is twofold. First, the spectral properties of the matrix model in the large $N$ limit do not depend on the details
of the matrix model potential, generic perturbations of the latter leave the former unchanged \cite{Akemann:1996vr, Damgaard:1997ye}. In (\ref{rmtz}) we have
chosen the simplest potential, a Gaussian. Second, the spectral properties of
two different underlying theories is expected to match the same matrix model
as long as $N_f$, rescaled fermion masses, number of zero modes and the pattern of chiral symmetry
breaking is left intact. The pattern of chiral symmetry breaking is a priori unknown, although the general expectation
\cite{Dimopoulos:1979sp} is that
since both the sextet and the fundamental representations are complex for $SU(3)$ the pattern will be the same, $SU(N_f)\times
SU(N_f) / SU(N_f)$. Hence it is expected that the same RMT ensemble, the chiral unitary, will describe both models.

The RMT predictions are different in each zero mode sector, labeled by $\nu$. For the fundamental representation the index
theorem equates $\nu$ with the topological charge but for higher dimensional representations the relation becomes $\nu = 2T Q$
where $T$ is the trace normalization factor of the representation. Overlap fermions possess exact zero modes giving rise to a
well-defined $\nu$ even at finite lattice spacing. However at finite lattice spacing $Q = \nu/2T$ can actually be a
fractional number but the number of configurations with this property disappears in the continuum limit \cite{Fodor:2009nh}. 

At realistic lattice spacing a sizable number of configurations have a non-integer $\nu/2T$ and
the distribution of Dirac eigenvalues on these configurations can also be compared with RMT. 
This is possible because the RMT
partition function is directly sensitive to the number of zero modes $\nu$ regardless of what its interpretation is in terms of
topological charges in the underlying theory.

In the quenched approximation, $N_f = 0$, the condition $m_\pi L \ll 1$ is not relevant but the first condition, $f_\pi L \gg 1$,
does impose a lower bound on $L$. How large $L$ has to be in order to separate the rotator and Goldstone modes in the finite
volume chiral Hamiltonian so that chiral perturbation theory is applicable, depends on the representation. 
We will see that different volumes will be required for the fundamental and sextet representations.
However once the volume is large enough, the universal behavior sets in and the choice of representation will not matter anymore.

\subsection{Conformal phase}

If the theory flows to a non-trivial conformal fixed point in the infrared, no scale is generated and $\Sigma = 0$. The
spectral density of the Dirac operator is
\bea
\rho(\lambda) \sim \lambda^{3-\gamma}\,,
\eea
where $\gamma$ is the anomalous dimension of ${\bar\psi} \psi$ and is calculable in perturbation theory. Up to 2-loops the universal
first two $\beta$-function coefficients determine the fixed point coupling $g_*$ and $\gamma = O(g_*^2)$. This calculation
can be trusted close to the upper end of the conformal window but might break down close to the lower end \cite{Banks:1981nn}.

In the free case $g_* = 0$ and the eigenvalue spacing is inversely proportional to the linear size $L$ in finite volume. The robust
prediction of RMT in the chirally broken case was that the eigenvalue spacing is inversely proportional to the volume $V = L^4$; see
(\ref{chideltal}). In
the non-trivial conformal case $g_* > 0$ and what is expected to hold is that the eigenvalue spacing is inversely proportional to the
linear size to some power which is much less than $4$ but of course larger than $1$. This is because anomalous dimensions can not
deviate too much from the free value because the fixed point $g_*$ can not deviate too much from $0$ since chiral symmetry
would be broken by a large coupling precluding the existence of a fixed point.

It is worth noting that the deconfined phase of QCD where chiral symmetry is restored, especially if the temperature is well
above $T_c$, might bare some resemblance to theories with a weak-coupling conformal fixed point. The eigenvalue distribution also
in this case does not follow RMT because chiral symmetry is intact. There exist models of the Dirac spectrum that match simulation
data rather nicely \cite{Kovacs:2008sc, Kovacs:2009zj}
and some ingredients of these models might very well prove useful for the study of weakly-coupled conformal gauge theories as well. 

\section{Simulations}

The exact solution of the RMT model in the large $N$ limit, given by (\ref{pzetak}, \ref{p12}) for the eigenvalue distributions, is quite
compact but still includes a number of integrals. The numerical evaluation of these integrals, especially for high $\nu$ and/or
$k$, is not
entirely straightforward due to large cancellations. Since in the sextet case $\nu=5Q$ one encounters $\nu = 10, 15, \ldots$ quite
frequently which are high enough so that the instabilities make the numerical integration cumbersome. We found it easier to
directly simulate the matrix model (\ref{rmtz}).

\subsection{RMT}

In each zero mode sector, $0 \leq \nu \leq 15$, the random matrix model (\ref{rmtz}) at $N_f = 0$ was simulated 
for $N = 300, 400, 500$ and $600$. The number of configurations was around $50,000$ for each and since their deviation from each
other was negligible relative to the statistical uncertainty in the gauge theory simulation, $N = 600$ was chosen
as the $N\to\infty$ limit. The RMT results quoted in the remainder of this work are based on $N = 600$ ensembles, with
about $110,000$ configurations in each sector (except for low $k$ and $\nu$ where the exact results are used in some of the
plots).

The expectation values of the first $10$ eigenvalues are listed in Table \ref{rmtaverages} for $0 \leq \nu \leq 15$.

\subsection{Gauge theory}

The eigenvalues of the overlap operator $D$ lie on a circle of radius $\rho/a$ centered at $(\rho/a, 0)$ where $-\rho/a$ is the
negative Wilson mass entering its definition. The non-real eigenvalues come in pairs, $\lambda_1 \pm i \lambda_2$, with
$\lambda_{1,2} > 0$. Picking the ones on the upper half plane, these are then
stereographically projected from $(2\rho/a,0)$ onto the positive imaginary axis to obtain the real and positive eigenvalues $\lambda$ 
that are used for the comparison with RMT,
\bea
\lambda = \frac{\lambda_2}{1 - \frac{a\lambda_1}{2\rho}}\;.
\eea

For the sextet representation, $2T = 5$. This means that about $5$ times more eigenvalues have to be computed for the sextet
operator than for the fundamental in order to end up with non-zero eigenvalues. The dimension of the sextet representation is $2$
times larger than the fundamental resulting in $4$ times more operations. These two facts result in about an order of magnitude
increase in computational cost for sextet eigenvalues relative to the fundamental. A comparison of RMT and QCD with overlap
fundamental fermions has been reported in \cite{Giusti:2003gf} for $N_f = 0$ and in \cite{Fukaya:2007yv} for $N_f = 2$. The sextet
representation and RMT have been considered in \cite{Damgaard:2001fg} within the staggered formulation.

The simulation parameters are summarized in Table \ref{simparam}. The lattices in set $A$ have physical volume $L=1.2215\,fm$ and
the lattices in set $B$ have $L=1.4658\,fm$. There are 5 and 2 lattice spacings in the two sets.
The scale is set by the static fundamental representation quark potential through $r_0 =
0.5\,fm$ \cite{Necco:2001xg}. The negative Wilson mass in the overlap operator was set to $-1.40$
for the fundamental and $-1.70$ for the sextet representation. Two steps of stout smearing was applied with smearing parameter
$0.15$ for both representations \cite{Morningstar:2003gk}. The implementation of the overlap operator is the same as in
\cite{Fodor:2003bh} and the gauge action is the Wilson plaquette.

\begin{table}
\begin{center}
\begin{tabular}{|cccccc|}
\hline
 &  $\beta$ & $L/a$ & $r_0/a$ & $a\;[fm]$  & \#config \\
\hline
\hline
$A_1$ & 5.9500  & 12  & 4.9122 & 0.1018 & 441 \\
\hline                                   
$A_2$ & 6.0384  & 14  & 5.7306 & 0.0873 & 543 \\
\hline                                   
$A_3$ & 6.1210  & 16  & 6.5496 & 0.0763 & 408 \\
\hline                                   
$A_4$ & 6.2719  & 20  & 8.1866 & 0.0611 & 211 \\
\hline                                   
$A_5$ & 6.4064  & 24  & 9.8239 & 0.0509 & 105 \\
\hline
\hline
$B_1$ & 6.0096  & 16  & 5.4574 & 0.0916 & 351 \\
\hline                                   
$B_2$ & 6.1474  & 20  & 6.8225 & 0.0733 & 228 \\
\hline
\end{tabular}
\caption{Simulation parameters for the Wilson plaquette gauge action. The physical volume is kept fixed in each set $A$ and $B$,
$L/r_0 = 2.4429$ and $2.9315$, respectively, corresponding to $L = 1.2215\,fm$ and $1.4658\,fm$,
respectively.}
\label{simparam}
\end{center}
\end{table}

\subsection{Parameter free comparisons with RMT}

\begin{figure}
\begin{center}
\begin{tabular}{c}
\includegraphics[width=7.5cm]{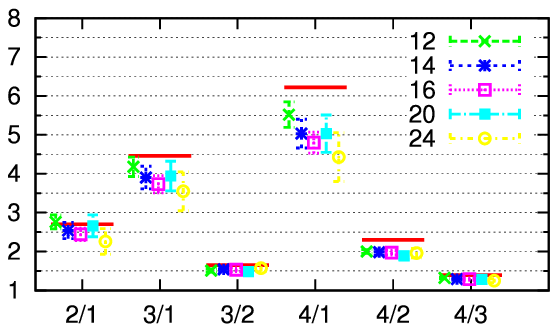} 
\includegraphics[width=7.5cm]{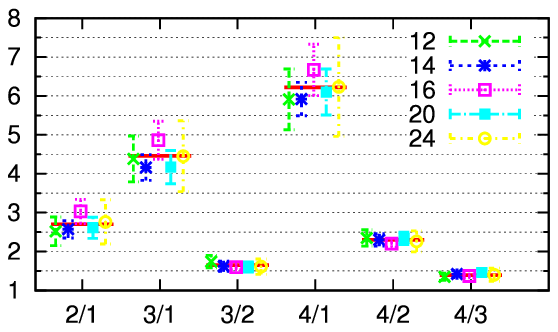} 
\end{tabular}
\caption{Ratios of measured average eigenvalues $\langle \lambda_i \rangle / \langle \lambda_j \rangle$ for the fundamental (left) and
sextet (right) representations on the $A$ lattices, $L=1.2215\,fm$, in the $\nu = 0$ sector for all five $L/a$ values. 
The RMT predictions (horizontal lines) are the same for both. \label{ratio1}}
\end{center}
\end{figure}

First let us compare the eigenvalue ratios $\langle \lambda_i \rangle / \langle \lambda_j \rangle$ with the RMT predictions. These
do not require the fitting of any parameter, these are completely parameter free predictions in each zero mode sector. 
Figure \ref{ratio1} shows these ratios for both representations for the $A$ lattices and $i,j = 1,2,3,4$ in the $\nu=0$ sector. 

Already from the $\nu = 0$ sector it is clear that the measured eigenvalue
ratios are consistently below the RMT predictions for the fundamental representation for most of the points, i.e. the fixed
$L=1.2215\,fm$ volume is too small. This same volume is large enough for the sextet representation though, and the measured
ratios fit the RMT predictions very nicely.

The observation that a given volume can be too small for the fundamental but large enough for the sextet representation is not
entirely a surprise. In \cite{Kogut:1984sb} it was noted long ago that the critical coupling of chiral symmetry breaking
$\beta_c$ for the sextet is much larger than for the
fundamental in the quenched approximation. This means that as long as $\beta$ is chosen the same and below
$\beta_c$ for both theories, the sextet model will be deeper in its chirally broken phase than the fundamental model.

\begin{figure}
\begin{center}
\begin{tabular}{c}
\includegraphics[width=7.5cm]{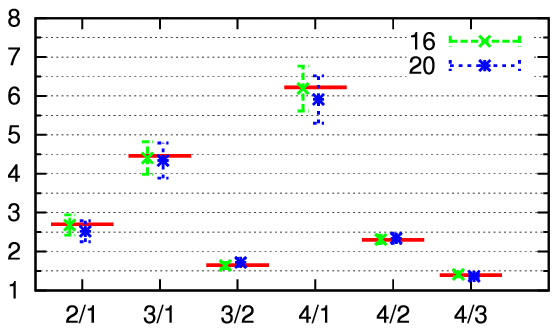}
\includegraphics[width=7.5cm]{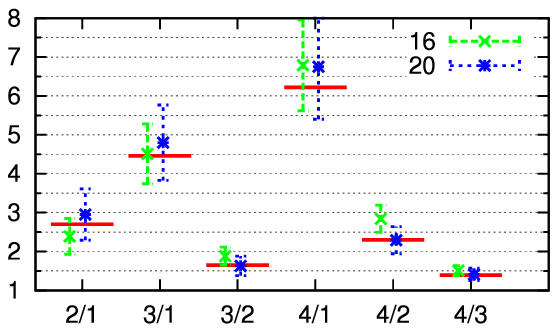} 
\end{tabular}
\caption{Ratios of measured average eigenvalues $\langle \lambda_i \rangle / \langle \lambda_j \rangle$ for the fundamental (left) and
sextet (right) representations on the $B$ lattices, $L=1.4658\,fm$, in the $\nu = 0$ sector for the two $L/a$ values. 
The RMT predictions (horizontal lines) are the same for both. \label{ratio2}}
\end{center}
\end{figure}

The eigenvalue ratios are shown on Figure \ref{ratio2} for the $B$ lattices where the volume is larger. This
volume is apparently large enough so that both theories enter the $\varepsilon$-regime and the RMT predictions work for both
representations. Since the pattern of symmetry breaking for the two representations is expected to be the same it is expected that the same
universal RMT describes the low-lying Dirac eigenvalues. Figure \ref{ratio2} confirms this universal property of RMT, the two
underlying theories are clearly different, but in the microscopic limit where RMT is expected to hold, they display the same
universal behavior.

Tables \ref{bigtable0} - \ref{bigtable15} list the eigenvalue ratios for all sectors $0 \leq \nu \leq 15$, both lattices $A$ and
$B$ and both representations (except for lattices $A$ and the fundamental where we have seen from the $\nu = 0$ sector that the
volume is too small). These tables are sparse for several reasons. One is that if the maximal fundamental index in an ensemble is
$\nu$, then the maximal sextet index will be roughly $5\nu$ and there will be no fundamental index between $\nu$ and $5\nu$. Also,
since the sextet index tends to come in 5-tuples for small lattice spacing, indices which are not a multiple of $5$ will have small or
no statistics. In addition, we calculate a fixed number of eigenvalues and it can happen that the number of computed 
eigenvalues is only 1, 2, 3, etc. more than the number of zero modes. In this case $\lambda_k$ for larger $k$ will not be available.

The eigenvalue ratios in the Tables \ref{bigtable0} - \ref{bigtable15} are mostly within one standard deviation from the RMT
predictions. More precisely, this happens for $65\%$ and $81\%$ of the cases for the sextet on the $B_1$ and $B_2$ lattices, 
$72\%$ and $53\%$ for the fundamental on the $B_1$ and $B_2$ lattices and $76\%, 86\%, 87\%, 87\%$ and $94\%$ of the cases on the
$A_1, \ldots, A_5$ lattices for the sextet. If the comparison is only made for the first $5$ eigenvalues then the results are 
within one standard deviation from the RMT predictions for $83\%, 91\%, 71\%, 78\%, 80\%, 85\%, 85\%, 87\%$ and $94\%$ of the
cases, in the same order as above. This we take as confirmation that the RMT predictions about the statistical properties of
small Dirac eigenvalues are correct.

\begin{figure}
\begin{center}
\begin{tabular}{ccc}
\includegraphics[width=4.8cm]{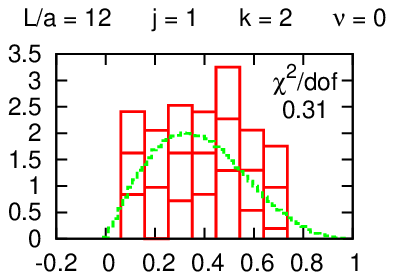} & \includegraphics[width=4.8cm]{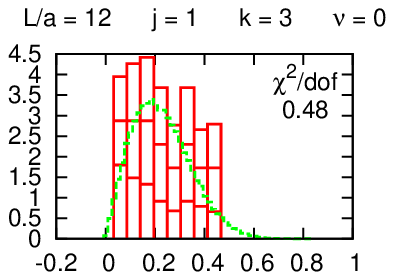} &
    \includegraphics[width=4.8cm]{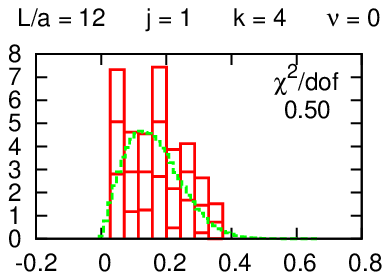} \\
\includegraphics[width=4.8cm]{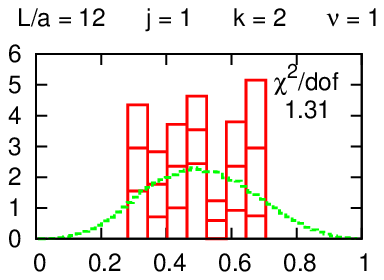} & \includegraphics[width=4.8cm]{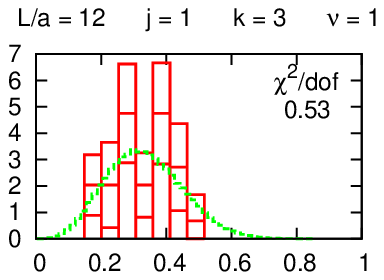} &
    \includegraphics[width=4.8cm]{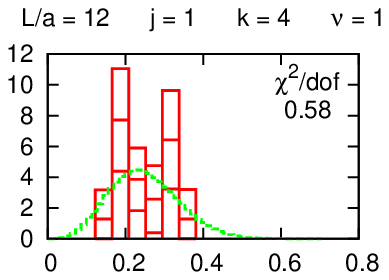} \\
\includegraphics[width=4.8cm]{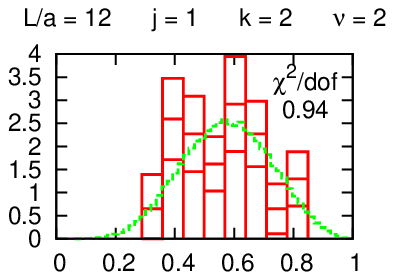} & \includegraphics[width=4.8cm]{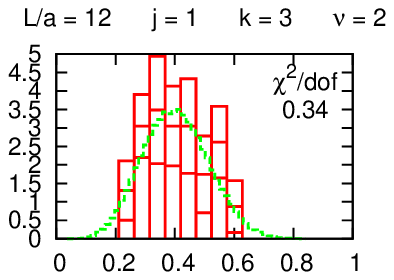} &
    \includegraphics[width=4.8cm]{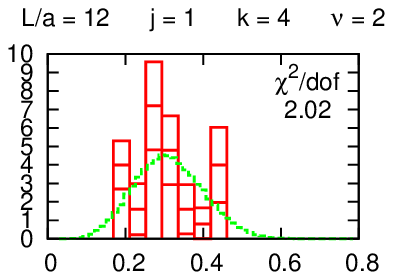} \\
\includegraphics[width=4.8cm]{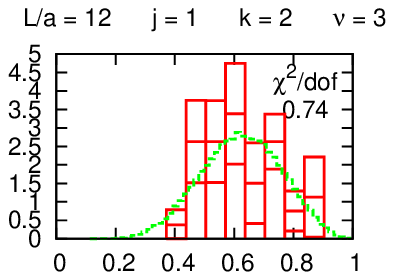} & \includegraphics[width=4.8cm]{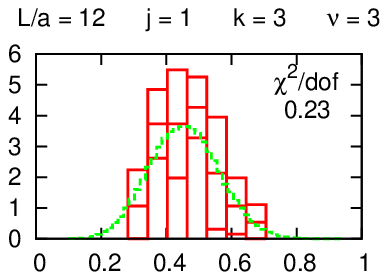} &
    \includegraphics[width=4.8cm]{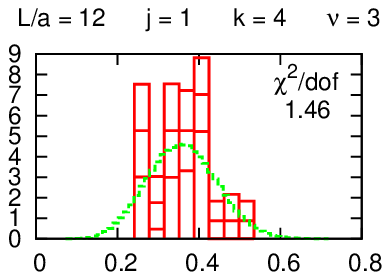} \\
\includegraphics[width=4.8cm]{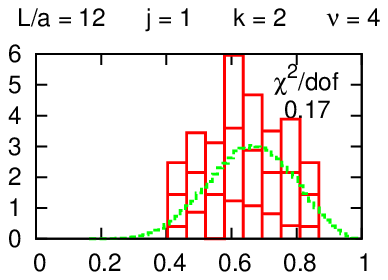} & \includegraphics[width=4.8cm]{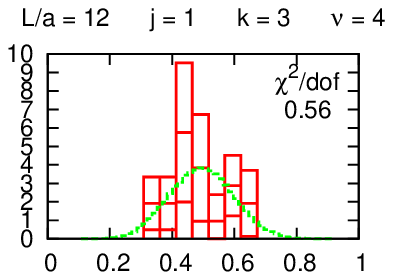} &
    \includegraphics[width=4.8cm]{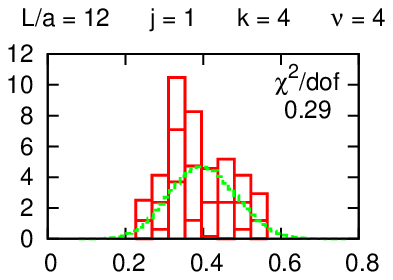} \\
\includegraphics[width=4.8cm]{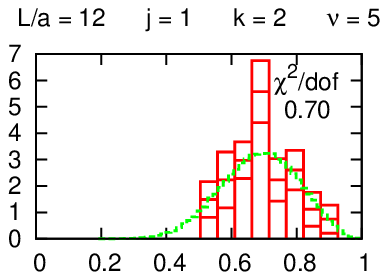} & \includegraphics[width=4.8cm]{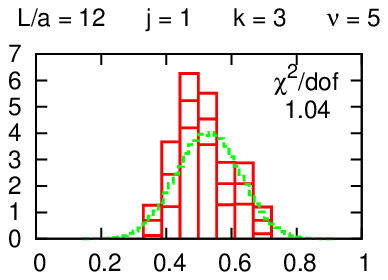} &
    \includegraphics[width=4.8cm]{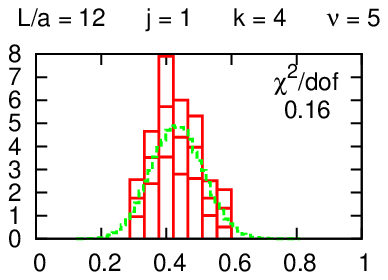} 
\end{tabular}
\caption{Distribution of the ratios $\lambda_j / \lambda_k$ in the sectors $\nu = 0, 1, 2, 3, 4, 5$ for the $A_1$ lattice and sextet
representation. The green (finer) histograms are from RMT, the red histograms are from the simulation including jackknife errors.\label{A1S}}
\end{center}
\end{figure}

\begin{figure}
\begin{center}
\begin{tabular}{ccc}
\includegraphics[width=4.8cm]{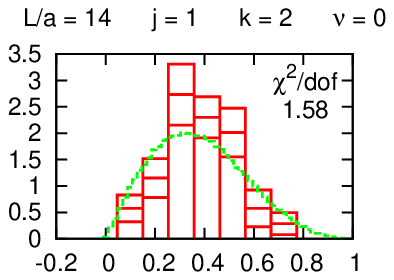} & \includegraphics[width=4.8cm]{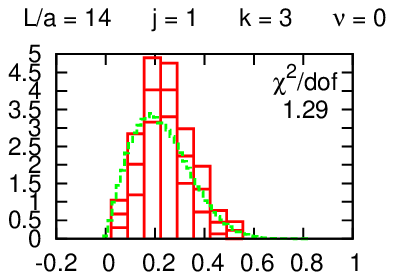} &
    \includegraphics[width=4.8cm]{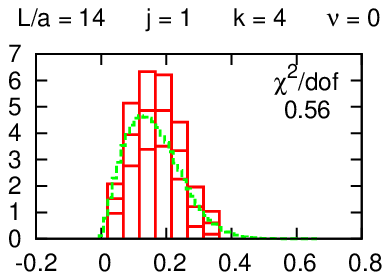} \\
\includegraphics[width=4.8cm]{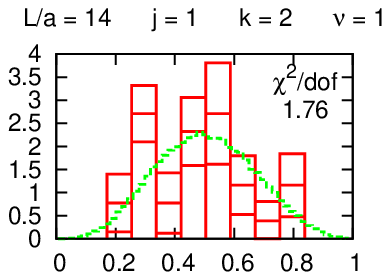} & \includegraphics[width=4.8cm]{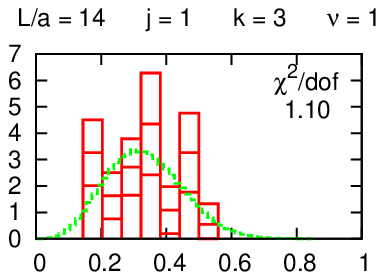} &
    \includegraphics[width=4.8cm]{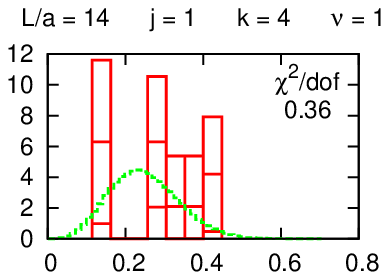} \\
\includegraphics[width=4.8cm]{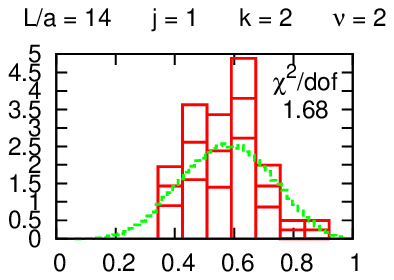} & \includegraphics[width=4.8cm]{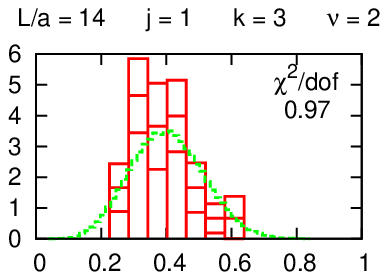} &
    \includegraphics[width=4.8cm]{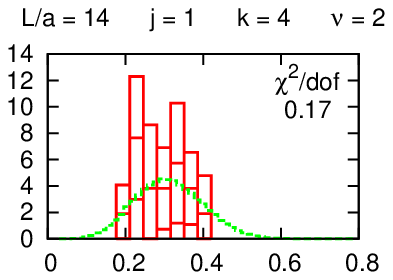} \\
\includegraphics[width=4.8cm]{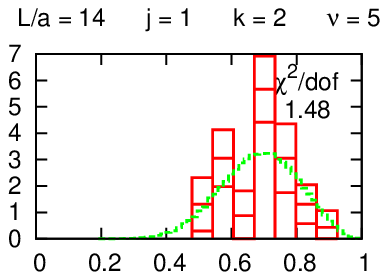} & \includegraphics[width=4.8cm]{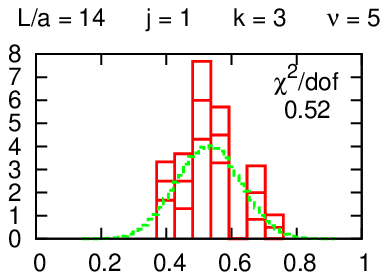} &
    \includegraphics[width=4.8cm]{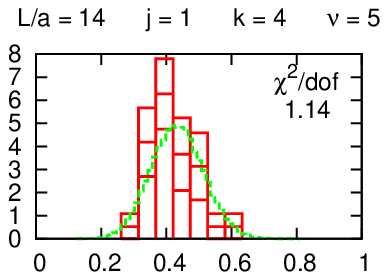} \\
\includegraphics[width=4.8cm]{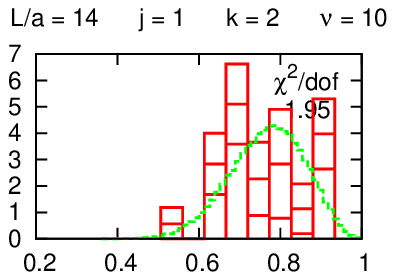} & \includegraphics[width=4.8cm]{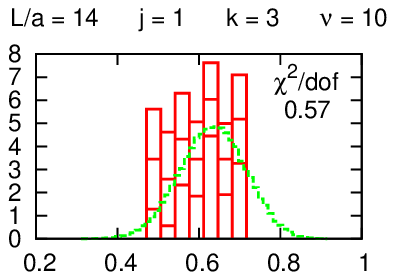} &
    \includegraphics[width=4.8cm]{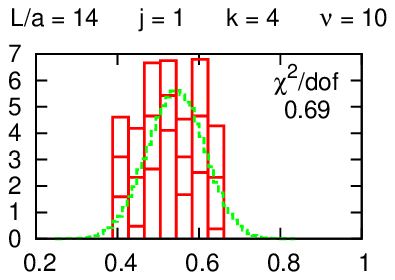} 
\end{tabular}
\caption{Distribution of the ratios $\lambda_j / \lambda_k$ in the sectors $\nu = 0, 1, 2, 5, 10$ for the $A_2$ lattice and sextet
representation. The green (finer) histograms are from RMT, the red histograms are from the simulation including jackknife errors.\label{A2S}}
\end{center}
\end{figure}

\begin{figure}
\begin{center}
\begin{tabular}{ccc}
\includegraphics[width=4.8cm]{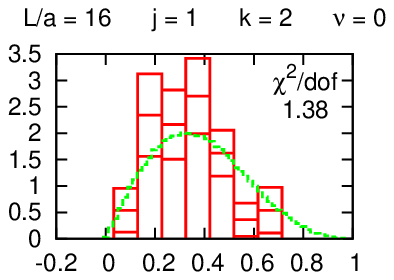} & \includegraphics[width=4.8cm]{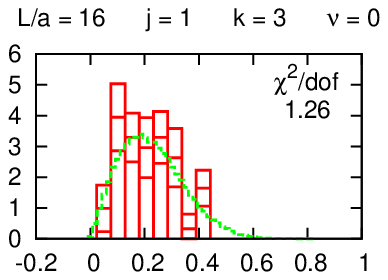} &
    \includegraphics[width=4.8cm]{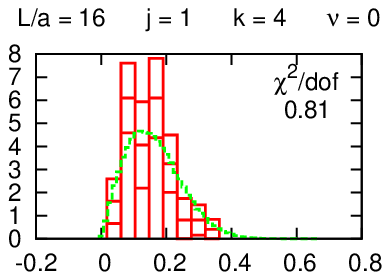} \\
\includegraphics[width=4.8cm]{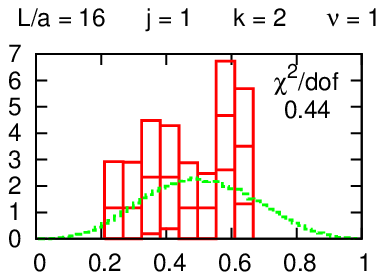} & \includegraphics[width=4.8cm]{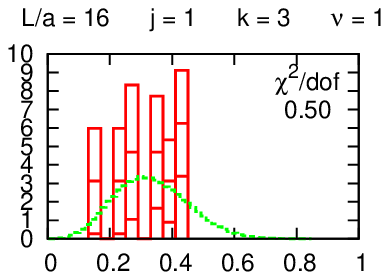} &
    \includegraphics[width=4.8cm]{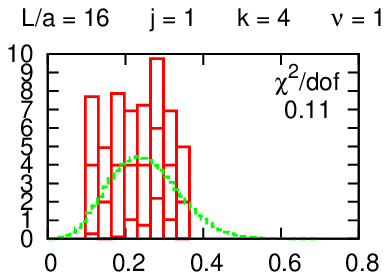} \\
\includegraphics[width=4.8cm]{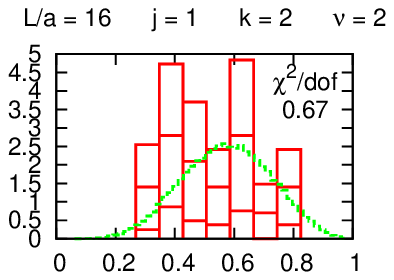} & \includegraphics[width=4.8cm]{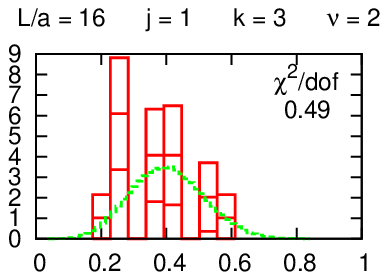} &
    \includegraphics[width=4.8cm]{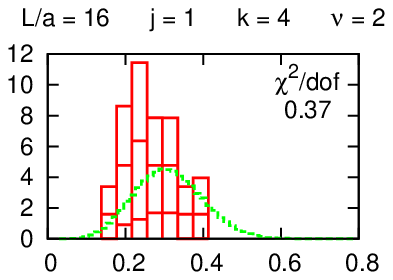} \\
\includegraphics[width=4.8cm]{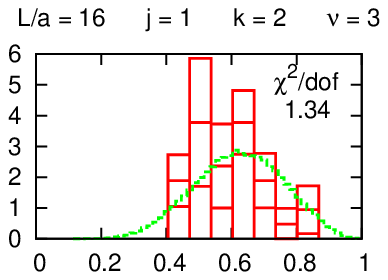} & \includegraphics[width=4.8cm]{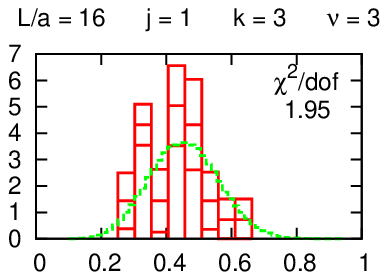} &
    \includegraphics[width=4.8cm]{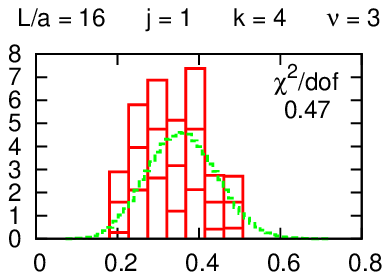} \\
\includegraphics[width=4.8cm]{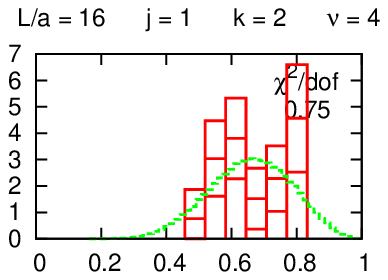} & \includegraphics[width=4.8cm]{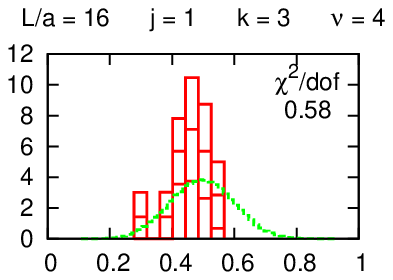} &
    \includegraphics[width=4.8cm]{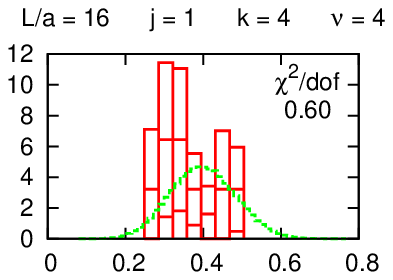} \\
\includegraphics[width=4.8cm]{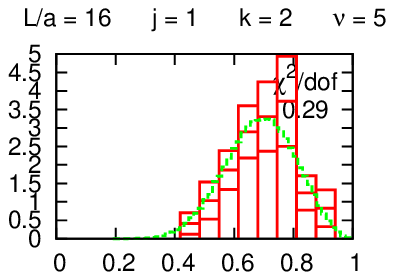} & \includegraphics[width=4.8cm]{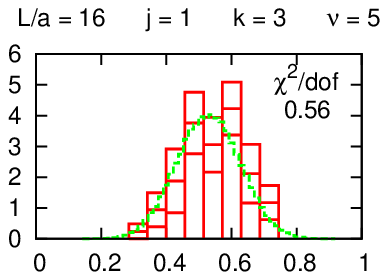} &
    \includegraphics[width=4.8cm]{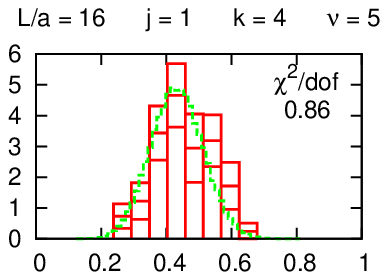} 
\end{tabular}
\caption{Distribution of the ratios $\lambda_j / \lambda_k$ in the sectors $\nu = 0, 1, 2, 3, 4, 5$ for the $A_3$ lattice and sextet
representation. The green (finer) histograms are from RMT, the red histograms are from the simulation including jackknife errors.\label{A3S}}
\end{center}
\end{figure}

\begin{figure}
\begin{center}
\begin{tabular}{ccc}
\includegraphics[width=4.8cm]{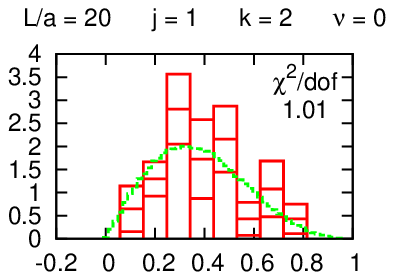} & \includegraphics[width=4.8cm]{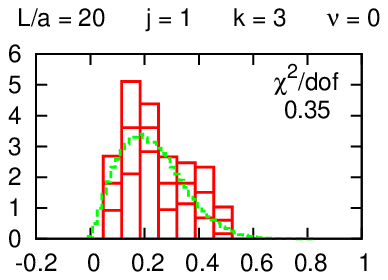} &
    \includegraphics[width=4.8cm]{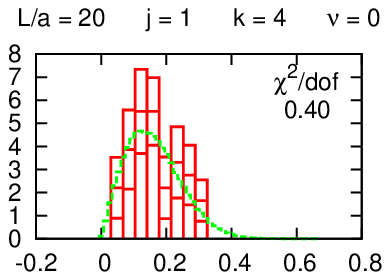} \\
\includegraphics[width=4.8cm]{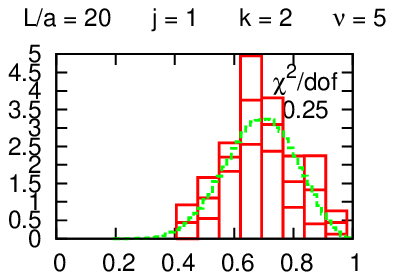} & \includegraphics[width=4.8cm]{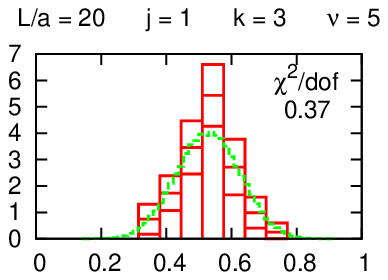} &
    \includegraphics[width=4.8cm]{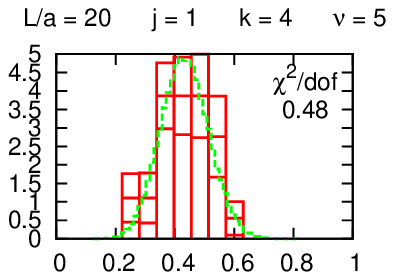} \\
\includegraphics[width=4.8cm]{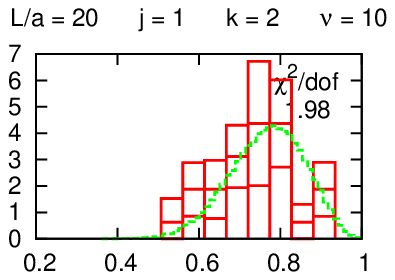} & \includegraphics[width=4.8cm]{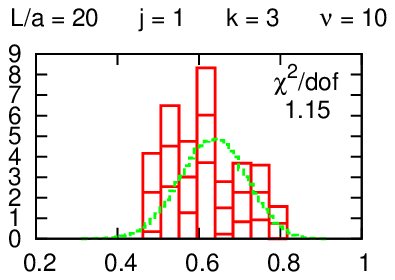} &
    \includegraphics[width=4.8cm]{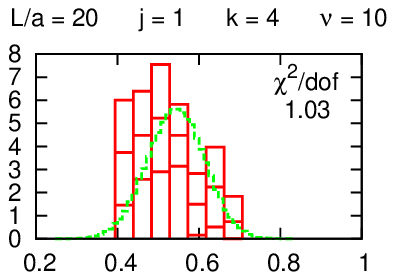} 
\end{tabular}
\caption{Distribution of the ratios $\lambda_j / \lambda_k$ in the sectors $\nu = 0, 5, 10$ for the $A_4$ lattice and sextet
representation. The green (finer) histograms are from RMT, the red histograms are from the simulation including jackknife errors.\label{A4S}}
\end{center}
\end{figure}

\begin{figure}
\begin{center}
\begin{tabular}{ccc}
\includegraphics[width=4.8cm]{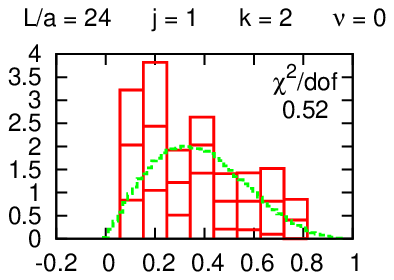} & \includegraphics[width=4.8cm]{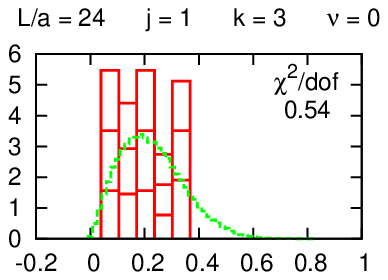} &
    \includegraphics[width=4.8cm]{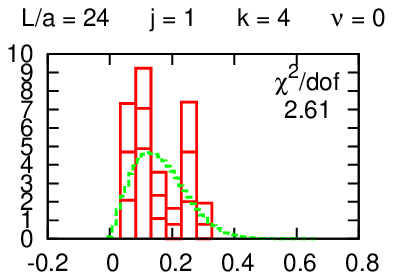} \\
\includegraphics[width=4.8cm]{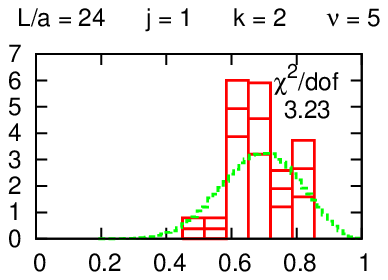} & \includegraphics[width=4.8cm]{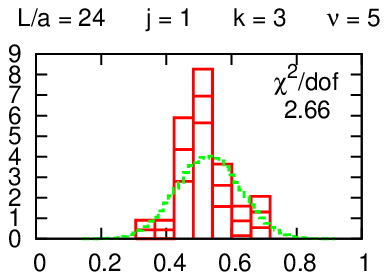} &
    \includegraphics[width=4.8cm]{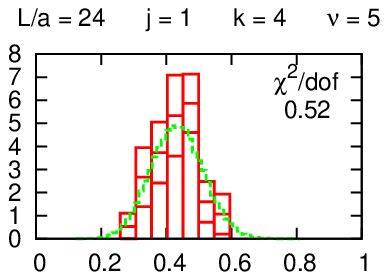} \\
\includegraphics[width=4.8cm]{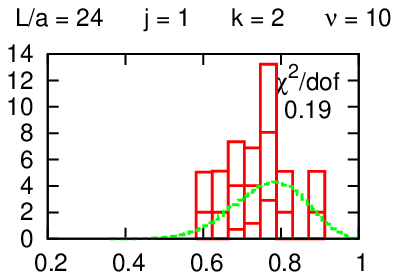} & \includegraphics[width=4.8cm]{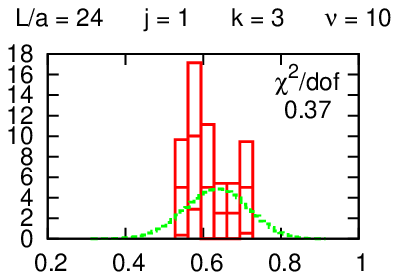} &
    \includegraphics[width=4.8cm]{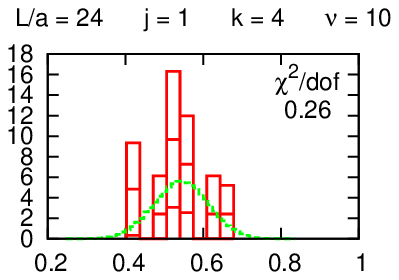}
\end{tabular}
\caption{Distribution of the ratios $\lambda_j / \lambda_k$ in the sectors $\nu = 0, 5, 10$ for the $A_5$ lattice and sextet
representation. The green (finer) histograms are from RMT, the red histograms are from the simulation including jackknife errors.\label{A5S}}
\end{center}
\end{figure}

\begin{figure}
\begin{center}
\begin{tabular}{ccc}
\includegraphics[width=4.8cm]{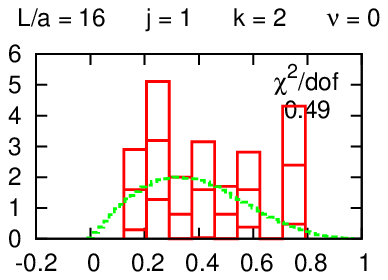} & \includegraphics[width=4.8cm]{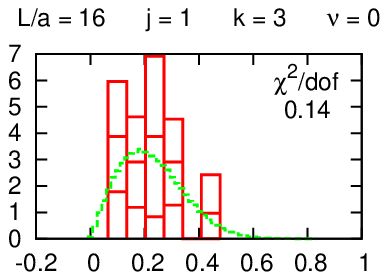} &
    \includegraphics[width=4.8cm]{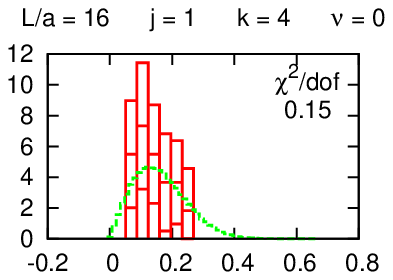} \\
\includegraphics[width=4.8cm]{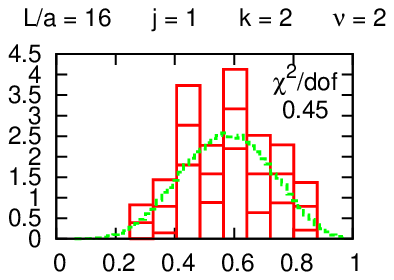} & \includegraphics[width=4.8cm]{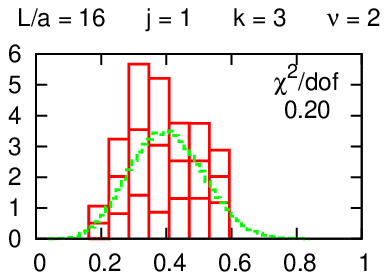} &
    \includegraphics[width=4.8cm]{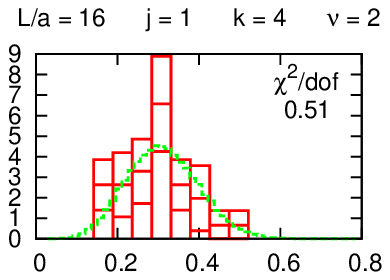} \\
\includegraphics[width=4.8cm]{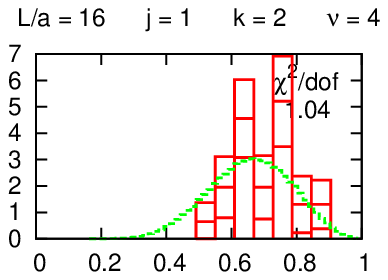} & \includegraphics[width=4.8cm]{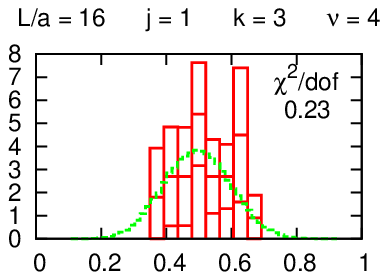} &
    \includegraphics[width=4.8cm]{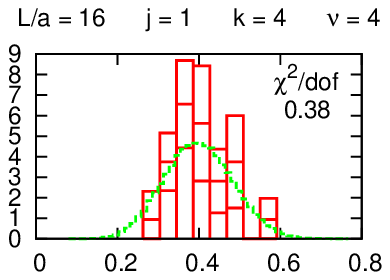} \\
\includegraphics[width=4.8cm]{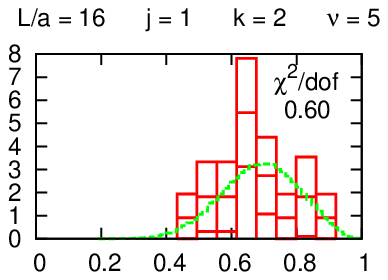} & \includegraphics[width=4.8cm]{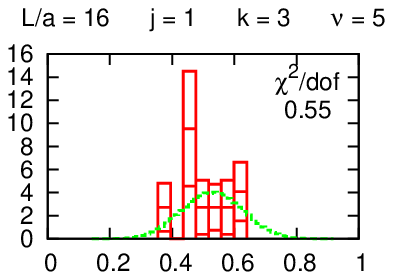} &
    \includegraphics[width=4.8cm]{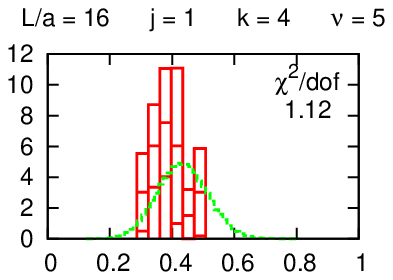} \\
\includegraphics[width=4.8cm]{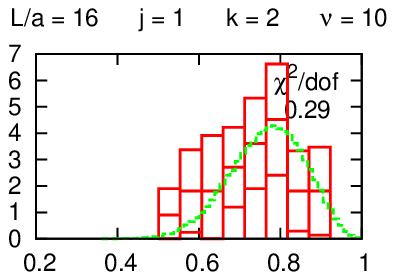} & \includegraphics[width=4.8cm]{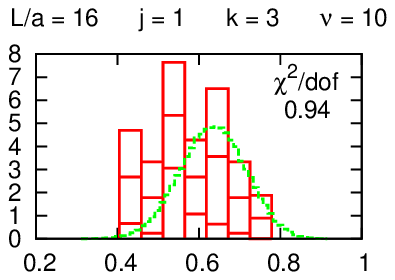} &
    \includegraphics[width=4.8cm]{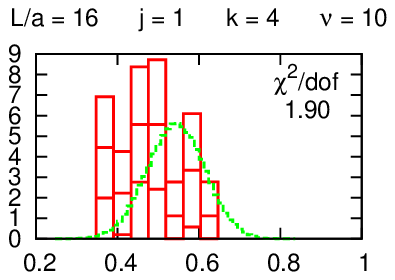} 
\end{tabular}
\caption{Distribution of the ratios $\lambda_j / \lambda_k$ in the sectors $\nu = 0, 2, 4, 5, 10$ for the $B_1$ lattice and sextet
representation. The green (finer) histograms are from RMT, the red histograms are from the simulation including jackknife errors.\label{B1S}}
\end{center}
\end{figure}

\begin{figure}
\begin{center}
\begin{tabular}{ccc}
\includegraphics[width=4.8cm]{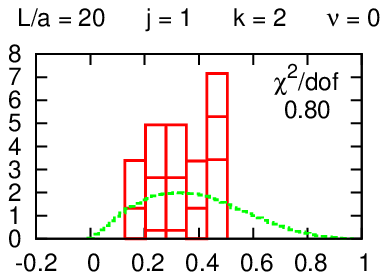} & \includegraphics[width=4.8cm]{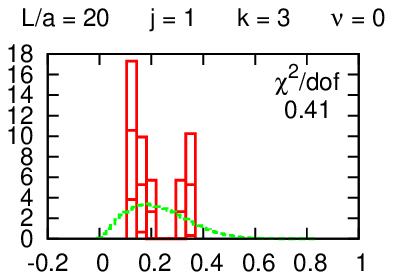} &
    \includegraphics[width=4.8cm]{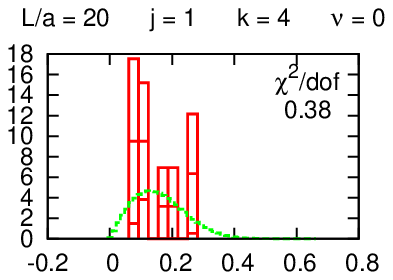} \\
\includegraphics[width=4.8cm]{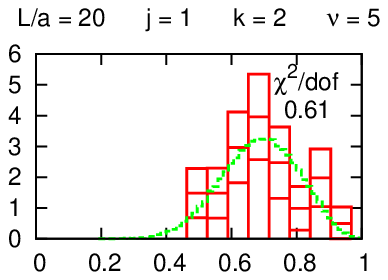} & \includegraphics[width=4.8cm]{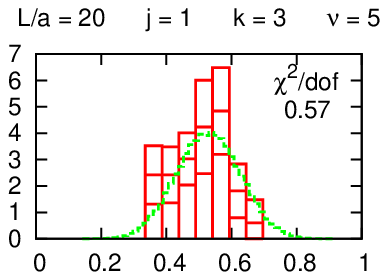} &
    \includegraphics[width=4.8cm]{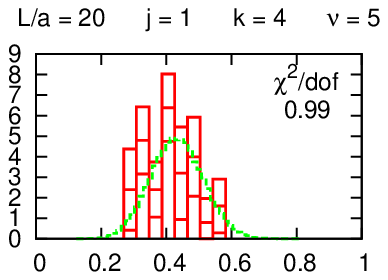} \\
\includegraphics[width=4.8cm]{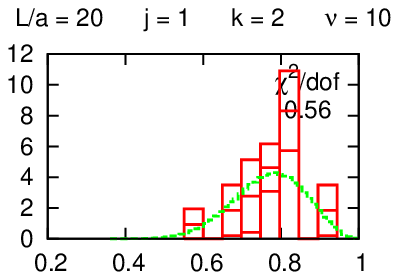} & \includegraphics[width=4.8cm]{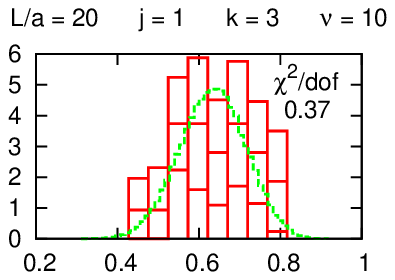} &
    \includegraphics[width=4.8cm]{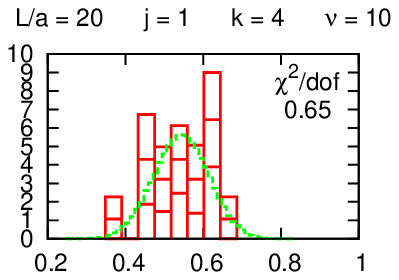} 
\end{tabular}
\caption{Distribution of the ratios $\lambda_j / \lambda_k$ in the sectors $\nu = 0, 5, 10$ for the $B_2$ lattice and sextet
representation. The green (finer) histograms are from RMT, the red histograms are from the simulation including jackknife errors.\label{B2S}}
\end{center}
\end{figure}

\begin{figure}
\begin{center}
\begin{tabular}{ccc}
\includegraphics[width=4.8cm]{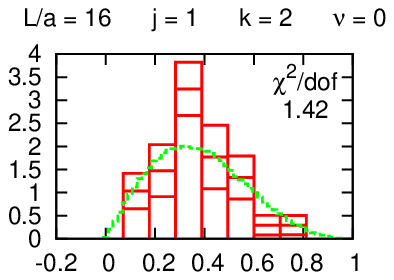} & \includegraphics[width=4.8cm]{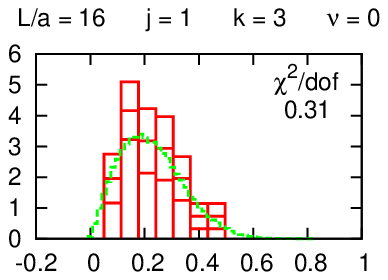} &
    \includegraphics[width=4.8cm]{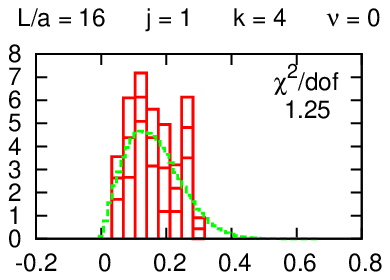} \\
\includegraphics[width=4.8cm]{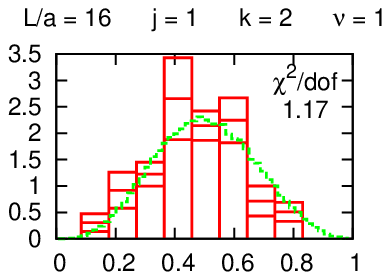} & \includegraphics[width=4.8cm]{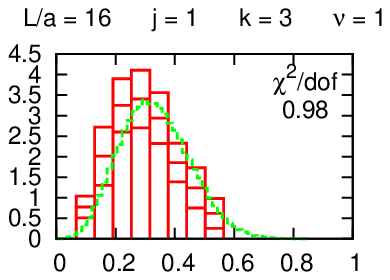} &
    \includegraphics[width=4.8cm]{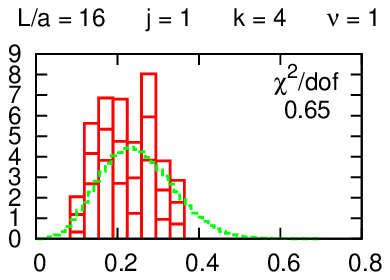} \\
\includegraphics[width=4.8cm]{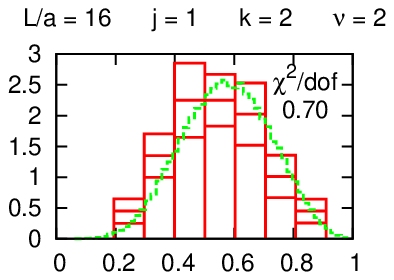} & \includegraphics[width=4.8cm]{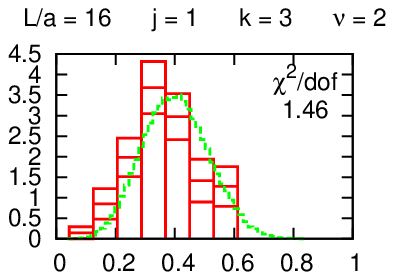} &
    \includegraphics[width=4.8cm]{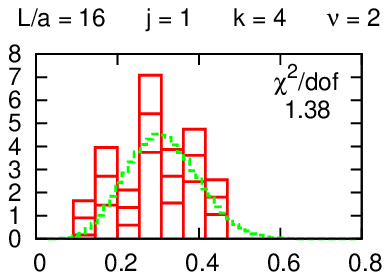} \\
\includegraphics[width=4.8cm]{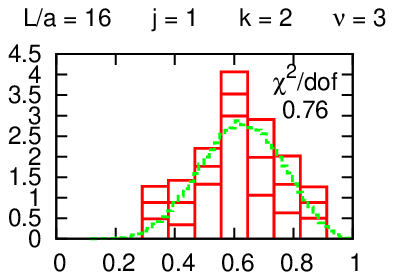} & \includegraphics[width=4.8cm]{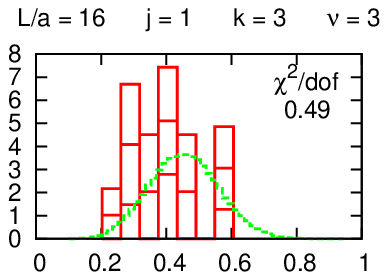} &
    \includegraphics[width=4.8cm]{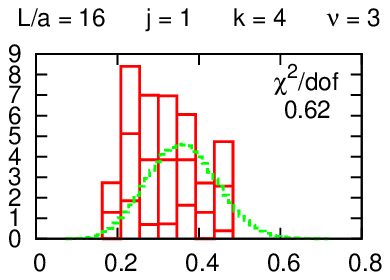}
\end{tabular}
\caption{Distribution of the ratios $\lambda_j / \lambda_k$ in the sectors $\nu = 0, 1, 2, 3$ for the $B_1$ lattice and fundamental
representation. The green (finer) histograms are from RMT, the red histograms are from the simulation including jackknife errors.\label{B1F}}
\end{center}
\end{figure}

\begin{figure}
\begin{center}
\begin{tabular}{ccc}
\includegraphics[width=4.8cm]{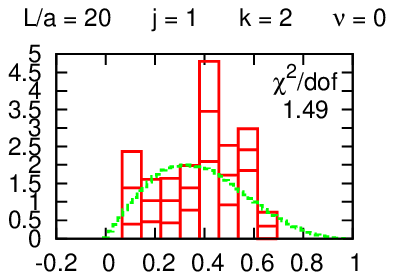} & \includegraphics[width=4.8cm]{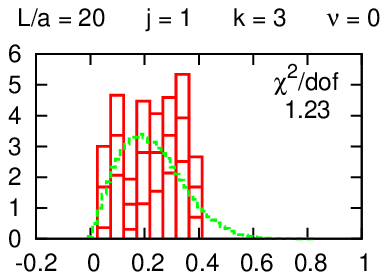} &
    \includegraphics[width=4.8cm]{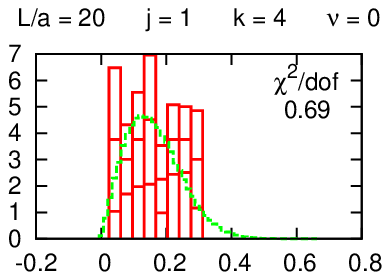} \\
\includegraphics[width=4.8cm]{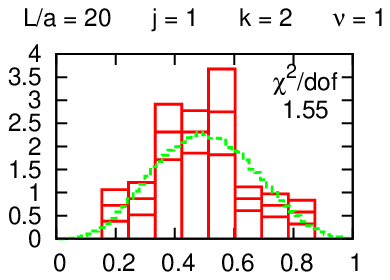} & \includegraphics[width=4.8cm]{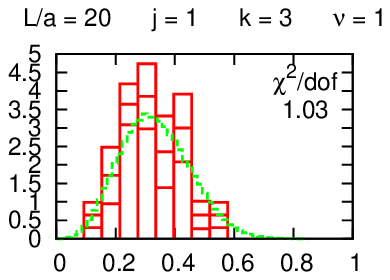} &
    \includegraphics[width=4.8cm]{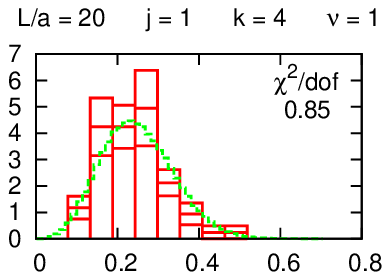} \\
\includegraphics[width=4.8cm]{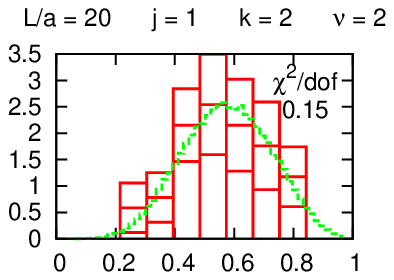} & \includegraphics[width=4.8cm]{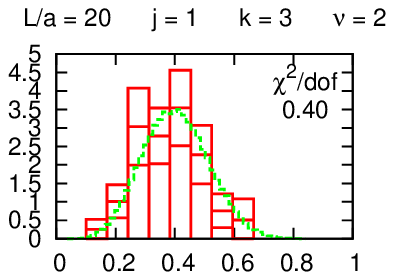} &
    \includegraphics[width=4.8cm]{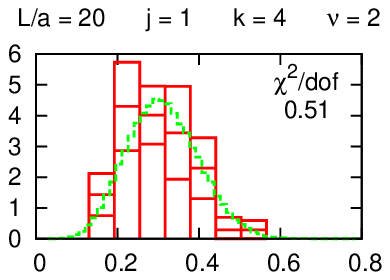} \\
\includegraphics[width=4.8cm]{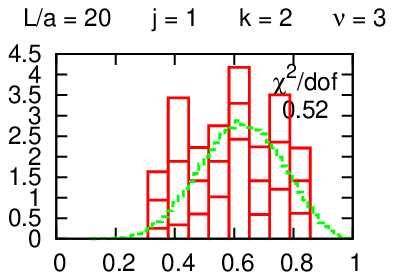} & \includegraphics[width=4.8cm]{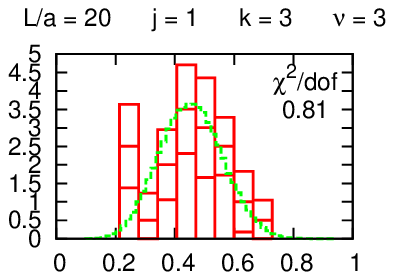} &
    \includegraphics[width=4.8cm]{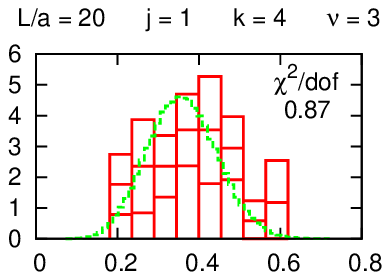} 
\end{tabular}
\caption{Distribution of the ratios $\lambda_j / \lambda_k$ in the sectors $\nu = 0, 1, 2, 3$ for the $B_2$ lattice and fundamental
representation. The green (finer) histograms are from RMT, the red histograms are from the simulation including jackknife errors.\label{B2F}}
\end{center}
\end{figure}

Not only the averages but also the entire distribution of eigenvalue ratios can be compared with the RMT predictions. Figures
\ref{A1S} - \ref{A5S} show the sextet eigenvalue ratio distributions in various zero mode sectors on the $A$ lattices. 
Figures \ref{B1S} and \ref{B2S} show the same for the larger $B$
lattices for the sextet, while fundamental eigenvalue ratio distributions are on Figures \ref{B1F} and \ref{B2F}.

The displayed errors on the histograms are jackknife errors. The level of agreement with the RMT
predictions may be characterized by the $\chi^2/dof$ value of the simulation results and the RMT curves using the jackknife errors
on each bin. These $\chi^2/dof$ values are displayed on each plot, they are all close to $1$. The smallest is $0.11$ and the largest is
$3.23$; for more details see the top right corner of each plot.

Clearly, the same quenched lattices induce the correct $\nu$-dependent distributions for the eigenvalue ratios depending on the
representation. The roughly 5-fold increase in $\nu$ for the sextet relative to the fundamental is correctly captured by the
$\nu$-dependent distribution of non-zero modes. These agreements are very non-trivial checks of the statement that the pattern of
chiral symmetry breaking is the same for the sextet as for the fundamental.

\subsection{Scale matching}

The only unknown parameter needed for a general comparison of simulations with RMT is the rescaling factor 
$\Sigma$, which in the quenched theory is only thought of as a low energy constant without a real physical meaning in terms of fermion bilinears.
In the previous section ratios of eigenvalues were considered from which $\Sigma$ drops out. In this section $\Sigma$
will be determined from the average smallest eigenvalue in the $\nu = 0$ sector and the predictions of RMT for eigenvalue
distributions in all sectors will be compared with the simulations.

This means that $\Sigma$ is determined from the relation,
\bea
\langle \zeta_1 \rangle_0 = \langle \lambda_1 \rangle_0 \Sigma V\;,
\eea
where the left hand side is evaluated from RMT and the right hand side is the result of the simulation. Since we have seen that
for the fundamental representation the $A$ lattices are too small, the comparison will not be done in this case. The resulting
values of $\Sigma$ are shown in Table \ref{sigmas}. As long as a not too high $k$ is chosen, a different determination of
$\Sigma$ from the $k^{th}$ eigenvalue in other zero mode sectors gives compatible values.

\begin{table}
\begin{center}
\begin{tabular}{|c|ccccc|cc|}
\hline
                 & $A_1$  & $A_2$  & $A_3$  & $A_4$  & $A_5$  & $B_1$   & $B_2$ \\
\hline
\hline
$a/r_0$          & 0.2036 & 0.1745 & 0.1527 & 0.1222 & 0.1018 & 0.1832  & 0.1466  \\
\hline
$r_0^3\Sigma_F$  &        &        &        &        &        & 0.37(6) & 0.35(3) \\
\hline
$r_0^3\Sigma_S$  & 2.0(2) & 2.2(1) & 2.5(3) & 2.5(2) & 2.6(5) & 2.9(5) & 2.9(3) \\
\hline
\end{tabular}
\caption{The chiral condensate $\Sigma$ from the average first eigenvalue in the $\nu=0$ sector for the two representations. 
The volume of the $A$ lattices is too small for the
fundamental representation and $\Sigma$ is not determined in this case. \label{sigmas}}
\end{center}
\end{table}

% The distribution of the first few eigenvalues is shown on Figures \ref{A1Sd} - \ref{A5Sd} for the $A$ lattices and sextet, on Figures 
% \ref{B1Sd} and \ref{B2Sd} for the $B$ lattices and sextet and finally on Figures \ref{B1Fd} and \ref{B2Fd} for the $B$ lattices
% and fundamental representation.

% Just as in the previous section some $\nu$ values are left out for some of the lattices in the sextet representation because those
% $\nu$ values which are not multiples of $5$ are suppressed. For the larger lattice spacings a considerable number of
% configurations exist where $\nu$ is not a multiple of $5$ and in these cases a comparison with RMT is meaningful, but as the
% lattice spacing decreases the number of such configurations becomes too low or even zero for any comparison.

Using the numerically determined chiral condensate the distributions of
the eigenvalues can be compared with the RMT predictions. The level of agreement is very similar to the case of the eigenvalue
ratios.

The non-trivial $\nu$-dependent matching \dash similarly to the parameter-free agreement of the previous section \dash between 
the gauge theory simulation in
the sextet representation and the chiral unitary RMT is showing that the pattern of chiral symmetry breaking is the same as for the fundamental
representation. It also suggests that if chiral symmetry breaking does take place with $N_f = 2$ sextet fermions then the pattern will also be the
same as with $N_f = 2$ fundamental fermions. If this happens to be the case then the number of Goldstone bosons will be $3$, as
expected on general grounds, which would be the appropriate number for phenomenological applications in technicolor models.
It should be noted that some, although non-conclusive, data has been presented in \cite{Shamir:2008pb, DeGrand:2009et} 
which appear to indicate that the $N_f = 2$ model might already be conformal.

\section{Conclusion}

We have shown that if the physical volume is large enough and the system enters into the $\varepsilon$-regime the predictions of
the same random matrix theory are accurately describing $SU(3)$ gauge theory with fermions in both the fundamental and sextet
representations at $N_f = 0$ and various zero mode sectors. Since the number of zero modes depends on the representation and the
matching with RMT is correct in this non-trivial $\nu$-dependent manner we take this as further evidence that the pattern of
chiral symmetry breaking is the same for the sextet and the fundamental, complementing earlier results with staggered fermions
in the trivial sector \cite{Damgaard:2001fg}.

An important aspect of the calculation was the use of overlap fermions which possess exact zero modes. In addition, it is expected
that for all gauge groups and all representations the universality class is correctly singled out by overlap fermions, a property
that is not present in the staggered formulation for certain real representations. These representations may turn out to play an
important role in technicolor models.

The application in technicolor models was our primary motivation. The topological and chiral properties of overlap fermions in 
higher dimensional representations can help in identifying in which phase a non-abelian gauge theory is in the infrared, whether
chiral symmetry breaking does take place or the dynamics is conformal. After the present quenched study we plan to extend these
methods to dynamical simulations where the potentially strong coupling can make the perturbative predictions regarding a conformal
fixed point unreliable.

\section*{Acknowledgments}

DN would like to acknowledge useful correspondence with Urs Heller.
The computations were carried out on the USQCD clusters at Fermilab and GPU clusters at Wuppertal University
\cite{Egri:2006zm}. This work is supported by the NSF under
grant 0704171, by the DOE under grants DOE-FG03-97ER40546, DOE-FG-02-97ER25308,
by the DFG under grant FO 502/1 and by SFB-TR/55.

\newpage

\begin{table}
\begin{center}
\begin{tabular}{|ccc|}
\hline
$\nu$   &       $k$     &   $\langle\zeta_k\rangle_\nu$ \\
\hline
\hline
0	&	1	&	1.777(6)	\\
\hline
0	&	2	&	4.79(1)	\\
\hline
0	&	3	&	7.90(2)	\\
\hline
0	&	4	&	11.03(3)	\\
\hline
0	&	5	&	14.16(4)	\\
\hline
0	&	6	&	17.30(5)	\\
\hline
0	&	7	&	20.44(6)	\\
\hline
0	&	8	&	23.57(7)	\\
\hline
0	&	9	&	26.71(8)	\\
\hline
0	&	10	&	29.85(9)	\\
\hline
\hline
1	&	1	&	3.11(1)	\\
\hline
1	&	2	&	6.26(2)	\\
\hline
1	&	3	&	9.42(3)	\\
\hline
1	&	4	&	12.56(4)	\\
\hline
1	&	5	&	15.69(5)	\\
\hline
1	&	6	&	18.83(6)	\\
\hline
1	&	7	&	21.98(6)	\\
\hline
1	&	8	&	25.12(7)	\\
\hline
1	&	9	&	28.26(8)	\\
\hline
1	&	10	&	31.41(9)	\\
\hline
\hline
2	&	1	&	4.34(1)	\\
\hline
2	&	2	&	7.64(2)	\\
\hline
2	&	3	&	10.84(3)	\\
\hline
2	&	4	&	14.01(4)	\\
\hline
2	&	5	&	17.17(5)	\\
\hline
2	&	6	&	20.33(6)	\\
\hline
2	&	7	&	23.48(7)	\\
\hline
2	&	8	&	26.62(8)	\\
\hline
2	&	9	&	29.77(9)	\\
\hline
2	&	10	&	32.92(10)	\\
\hline
\hline
3	&	1	&	5.53(2)	\\
\hline
3	&	2	&	8.95(3)	\\
\hline
3	&	3	&	12.22(4)	\\
\hline
3	&	4	&	15.42(5)	\\
\hline
3	&	5	&	18.61(6)	\\
\hline
3	&	6	&	21.78(6)	\\
\hline
3	&	7	&	24.93(7)	\\
\hline
3	&	8	&	28.09(8)	\\
\hline
3	&	9	&	31.24(9)	\\
\hline
3	&	10	&	34.39(10)	\\
\hline
\end{tabular}
\begin{tabular}{|ccc|}
\hline
$\nu$   &       $k$     &   $\langle\zeta_k\rangle_\nu$ \\
\hline
\hline
4	&	1	&	6.68(2)	\\
\hline
4	&	2	&	10.23(3)	\\
\hline
4	&	3	&	13.55(4)	\\
\hline
4	&	4	&	16.80(5)	\\
\hline
4	&	5	&	20.00(6)	\\
\hline
4	&	6	&	23.19(7)	\\
\hline
4	&	7	&	26.37(8)	\\
\hline
4	&	8	&	29.55(9)	\\
\hline
4	&	9	&	32.70(10)	\\
\hline
4	&	10	&	35.86(11)	\\
\hline
\hline
5	&	1	&	7.83(2)	\\
\hline
5	&	2	&	11.47(3)	\\
\hline
5	&	3	&	14.86(4)	\\
\hline
5	&	4	&	18.14(5)	\\
\hline
5	&	5	&	21.38(6)	\\
\hline
5	&	6	&	24.59(7)	\\
\hline
5	&	7	&	27.78(8)	\\
\hline
5	&	8	&	30.96(9)	\\
\hline
5	&	9	&	34.13(10)	\\
\hline
5	&	10	&	37.29(11)	\\
\hline
\hline
6	&	1	&	8.94(3)	\\
\hline
6	&	2	&	12.69(4)	\\
\hline
6	&	3	&	16.14(5)	\\
\hline
6	&	4	&	19.46(6)	\\
\hline
6	&	5	&	22.73(7)	\\
\hline
6	&	6	&	25.97(8)	\\
\hline
6	&	7	&	29.17(9)	\\
\hline
6	&	8	&	32.36(10)	\\
\hline
6	&	9	&	35.55(11)	\\
\hline
6	&	10	&	38.73(12)	\\
\hline
\hline
7	&	1	&	10.04(3)	\\
\hline
7	&	2	&	13.90(4)	\\
\hline
7	&	3	&	17.39(5)	\\
\hline
7	&	4	&	20.75(6)	\\
\hline
7	&	5	&	24.05(7)	\\
\hline
7	&	6	&	27.31(8)	\\
\hline
7	&	7	&	30.54(9)	\\
\hline
7	&	8	&	33.74(10)	\\
\hline
7	&	9	&	36.94(11)	\\
\hline
7	&	10	&	40.12(12)	\\
\hline
\end{tabular}
\begin{tabular}{|ccc|}
\hline
$\nu$   &       $k$     &   $\langle\zeta_k\rangle_\nu$ \\
\hline
\hline
8	&	1	&	11.15(3)	\\
\hline
8	&	2	&	15.09(4)	\\
\hline
8	&	3	&	18.64(6)	\\
\hline
8	&	4	&	22.04(7)	\\
\hline
8	&	5	&	25.36(8)	\\
\hline
8	&	6	&	28.64(8)	\\
\hline
8	&	7	&	31.89(9)	\\
\hline
8	&	8	&	35.11(10)	\\
\hline
8	&	9	&	38.32(11)	\\
\hline
8	&	10	&	41.52(12)	\\
\hline
\hline
9	&	1	&	12.24(4)	\\
\hline
9	&	2	&	16.26(5)	\\
\hline
9	&	3	&	19.87(6)	\\
\hline
9	&	4	&	23.31(7)	\\
\hline
9	&	5	&	26.67(8)	\\
\hline
9	&	6	&	29.97(9)	\\
\hline
9	&	7	&	33.23(10)	\\
\hline
9	&	8	&	36.47(11)	\\
\hline
9	&	9	&	39.69(12)	\\
\hline
9	&	10	&	42.89(13)	\\
\hline
\hline
10	&	1	&	13.32(4)	\\
\hline
10	&	2	&	17.42(5)	\\
\hline
10	&	3	&	21.08(6)	\\
\hline
10	&	4	&	24.56(7)	\\
\hline
10	&	5	&	27.95(8)	\\
\hline
10	&	6	&	31.26(9)	\\
\hline
10	&	7	&	34.54(10)	\\
\hline
10	&	8	&	37.79(11)	\\
\hline
10	&	9	&	41.03(12)	\\
\hline
10	&	10	&	44.25(13)	\\
\hline
\hline
11	&	1	&	14.39(4)	\\
\hline
11	&	2	&	18.59(6)	\\
\hline
11	&	3	&	22.28(7)	\\
\hline
11	&	4	&	25.80(8)	\\
\hline
11	&	5	&	29.20(9)	\\
\hline
11	&	6	&	32.55(10)	\\
\hline
11	&	7	&	35.85(11)	\\
\hline
11	&	8	&	39.12(12)	\\
\hline
11	&	9	&	42.37(13)	\\
\hline
11	&	10	&	45.60(14)	\\
\hline
\end{tabular}
\begin{tabular}{|ccc|}
\hline
$\nu$   &       $k$     &   $\langle\zeta_k\rangle_\nu$ \\
\hline
\hline
12	&	1	&	15.47(5)	\\
\hline
12	&	2	&	19.73(6)	\\
\hline
12	&	3	&	23.48(7)	\\
\hline
12	&	4	&	27.02(8)	\\
\hline
12	&	5	&	30.46(9)	\\
\hline
12	&	6	&	33.83(10)	\\
\hline
12	&	7	&	37.14(11)	\\
\hline
12	&	8	&	40.44(12)	\\
\hline
12	&	9	&	43.70(13)	\\
\hline
12	&	10	&	46.93(14)	\\
\hline
\hline
13	&	1	&	16.53(5)	\\
\hline
13	&	2	&	20.86(6)	\\
\hline
13	&	3	&	24.66(7)	\\
\hline
13	&	4	&	28.24(8)	\\
\hline
13	&	5	&	31.71(10)	\\
\hline
13	&	6	&	35.09(11)	\\
\hline
13	&	7	&	38.43(12)	\\
\hline
13	&	8	&	41.73(13)	\\
\hline
13	&	9	&	45.01(13)	\\
\hline
13	&	10	&	48.26(14)	\\
\hline
\hline
14	&	1	&	17.60(5)	\\
\hline
14	&	2	&	22.00(7)	\\
\hline
14	&	3	&	25.84(8)	\\
\hline
14	&	4	&	29.45(9)	\\
\hline
14	&	5	&	32.95(10)	\\
\hline
14	&	6	&	36.36(11)	\\
\hline
14	&	7	&	39.72(12)	\\
\hline
14	&	8	&	43.03(13)	\\
\hline
14	&	9	&	46.33(14)	\\
\hline
14	&	10	&	49.59(15)	\\
\hline
\hline
15	&	1	&	18.65(6)	\\
\hline
15	&	2	&	23.13(7)	\\
\hline
15	&	3	&	27.01(8)	\\
\hline
15	&	4	&	30.66(9)	\\
\hline
15	&	5	&	34.18(10)	\\
\hline
15	&	6	&	37.61(11)	\\
\hline
15	&	7	&	40.98(12)	\\
\hline
15	&	8	&	44.32(13)	\\
\hline
15	&	9	&	47.62(14)	\\
\hline
15	&	10	&	50.89(15)	\\
\hline
\end{tabular}
\caption{RMT expectation values of the first 10 eigenvalues in each sector $\nu$.\label{rmtaverages}}
\end{center}
\end{table}

\newpage

\begin{table}
\small
\begin{center}
\begin{tabular}{|ccccccccccc|}
\hline
$j/k$ & RMT & $B_1(S)$ & $B_2(S)$ & $B_1(F)$ & $B_2(F)$ & $A_1(S)$ & $A_2(S)$ & $A_3(S)$ & $A_4(S)$ & $A_5(S)$ \\
\hline
\hline
2/1&2.703&2.39(46)&2.95(66)&2.68(26)&2.52(27)&2.52(37)&2.57(22)&3.03(31)&2.61(27)&2.76(57) \\
\hline
3/1&4.460&4.51(77)&4.80(97)&4.40(42)&4.34(45)&4.38(59)&4.16(33)&4.86(49)&4.17(43)&4.45(91) \\
\hline
4/1&6.220&6.8(1.2)&6.8(1.4)&6.19(58)&5.91(61)&5.91(78)&5.92(43)&6.67(66)&6.10(59)&6.2(1.3) \\
\hline
5/1&7.991&8.7(1.5)&8.1(1.6)&7.99(75)&7.30(74)&7.12(94)&7.87(71)&8.66(86)&7.73(72)&8.2(1.6) \\
\hline
6/1&9.759&10.4(1.8)&10.1(1.9)&9.49(87)&8.69(87)&8.8(1.2)&9.45(88)&10.6(1.0)&9.55(89)&10.0(1.9) \\
\hline
7/1&11.532&12.6(2.1)&12.0(2.3)& &9.54(99)& &11.18(94)&13.0(1.3)&11.3(1.0)&11.4(2.2) \\
\hline
8/1&13.300&14.4(2.5)&14.3(2.7)& & & &13.1(1.1)&14.9(1.6)&12.9(1.2)&13.1(2.5) \\
\hline
\end{tabular}
\caption{Average eigenvalue ratios in the $\nu = 0$ sector in RMT and the $A_i$ and $B_i$ lattices.}\label{bigtable0}
\end{center}
\end{table}

\newpage

\begin{table}
\small
\begin{center}
\begin{tabular}{|ccccccccccc|}
\hline
$j/k$ & RMT & $B_1(S)$ & $B_2(S)$ & $B_1(F)$ & $B_2(F)$ & $A_1(S)$ & $A_2(S)$ & $A_3(S)$ & $A_4(S)$ & $A_5(S)$ \\
\hline
\hline
2/1&2.016&1.95(20)&1.64(23)& 2.20(9)&2.10(10)&2.01(14)&2.09(24)&2.07(34)&2.08(52)&  \\
\hline
3/1&3.031&2.87(28)&2.54(38)&3.37(13)&3.15(16)&3.05(24)&3.00(32)&3.18(51)&3.35(78)&  \\
\hline
4/1&4.041&3.79(39)&3.22(44)&4.63(17)&4.18(19)&3.97(35)&4.13(56)&4.30(68)&4.9(1.1)&  \\
\hline
5/1&5.050&4.80(52)&4.21(58)&5.67(21)&5.12(23)&5.07(46)&5.41(75)&5.46(87)&6.5(1.5)&  \\
\hline
6/1&6.060&5.73(59)&5.23(72)&6.55(24)&5.95(26)&6.02(53)&6.51(89)&6.9(1.1)&8.2(1.9)&  \\
\hline
7/1&7.073&6.79(70)&6.02(83)&7.5(1.1)&6.59(29)& &7.38(97)&8.2(1.3)&9.4(2.2)&  \\
\hline
8/1&8.082&8.14(86)&6.96(97)& & & &8.7(1.1)&9.5(1.4)&10.7(2.5)&  \\
\hline
\end{tabular}
\caption{Average eigenvalue ratios in the $\nu = 1$ sector in RMT and the $A_i$ and $B_i$ lattices.}\label{bigtable1}
\end{center}
\end{table}

\newpage

\begin{table}
\small
\begin{center}
\begin{tabular}{|ccccccccccc|}
\hline
$j/k$ & RMT & $B_1(S)$ & $B_2(S)$ & $B_1(F)$ & $B_2(F)$ & $A_1(S)$ & $A_2(S)$ & $A_3(S)$ & $A_4(S)$ & $A_5(S)$ \\
\hline
\hline
2/1&1.761&1.73(14)&1.55(42)& 1.85(9)&1.82(15)&1.79(14)&1.77(11)&1.96(13)&1.87(40)&1.57(78) \\
\hline
3/1&2.500&2.53(21)&2.32(57)&2.75(13)&2.60(20)&2.49(18)&2.54(16)&2.89(24)&2.82(56)&2.4(1.2) \\
\hline
4/1&3.239&3.33(24)&2.96(70)&3.65(18)&3.33(25)&3.30(25)&3.68(31)&3.75(26)&3.35(67)&3.1(1.5) \\
\hline
5/1&3.953&3.99(28)&3.77(85)&4.37(21)&4.05(30)&4.11(30)&4.42(33)&4.71(32)&4.34(88)&4.0(1.9) \\
\hline
6/1&4.680&4.80(34)&4.5(1.0)& &4.59(34)& &5.19(38)&5.38(51)&5.3(1.1)&5.0(2.3) \\
\hline
7/1&5.404&5.61(40)&5.3(1.2)& & & &6.06(40)&6.44(61)&6.3(1.2)&5.8(2.6) \\
\hline
8/1&6.129&6.45(46)&6.1(1.4)& & & &6.78(42)&7.44(77)&7.0(1.3)&6.5(3.0) \\
\hline
\end{tabular}
\caption{Average eigenvalue ratios in the $\nu = 2$ sector in RMT and the $A_i$ and $B_i$ lattices.}\label{bigtable2}
\end{center}
\end{table}

\newpage

\begin{table}
\small
\begin{center}
\begin{tabular}{|ccccccccccc|}
\hline
$j/k$ & RMT & $B_1(S)$ & $B_2(S)$ & $B_1(F)$ & $B_2(F)$ & $A_1(S)$ & $A_2(S)$ & $A_3(S)$ & $A_4(S)$ & $A_5(S)$ \\
\hline
\hline
2/1&1.617&1.63(15)&1.73(18)&1.68(11)&1.67(11)&1.61(10)&1.60(11)&1.67(13)&2.31(62)& 1.49(8) \\
\hline
3/1&2.208&2.25(22)&2.32(20)&2.30(15)&2.20(12)&2.15(13)&2.23(17)&2.35(18)&3.57(75)&1.93(18) \\
\hline
4/1&2.787&2.82(28)&2.92(25)&2.91(19)&2.61(15)&2.79(16)&2.88(21)&3.09(22)&4.40(84)&2.62(11) \\
\hline
5/1&3.363&3.44(33)&3.51(31)&3.34(20)&3.03(17)&3.41(20)&3.47(27)&3.73(27)&5.9(1.1)&3.01(13) \\
\hline
6/1&3.936&4.03(40)&4.16(32)& &3.43(19)& &4.17(31)&4.32(32)&6.7(1.3)&3.74(21) \\
\hline
7/1&4.506&4.70(46)&4.83(39)& & & &4.86(36)&4.94(39)&7.4(1.4)&4.26(25) \\
\hline
8/1&5.077&5.35(50)&5.45(44)& & & &5.48(40)&5.59(41)& &4.87(41) \\
\hline
\end{tabular}
\caption{Average eigenvalue ratios in the $\nu = 3$ sector in RMT and the $A_i$ and $B_i$ lattices.}\label{bigtable3}
\end{center}
\end{table}

\newpage

\begin{table}
\small
\begin{center}
\begin{tabular}{|ccccccccccc|}
\hline
$j/k$ & RMT & $B_1(S)$ & $B_2(S)$ & $B_1(F)$ & $B_2(F)$ & $A_1(S)$ & $A_2(S)$ & $A_3(S)$ & $A_4(S)$ & $A_5(S)$ \\
\hline
\hline
2/1&1.529& 1.41(7)&1.35(15)&1.48(12)&1.52(12)&1.58(21)&1.67(15)&1.51(12)&1.39(32)&  \\
\hline
3/1&2.025& 1.91(9)&1.96(25)&1.99(16)&1.97(14)&2.07(27)&1.89(21)&2.09(15)&1.89(45)&  \\
\hline
4/1&2.510&2.46(11)&2.39(31)&2.50(17)&2.38(18)&2.58(33)&2.40(29)&2.58(19)&2.33(51)&  \\
\hline
5/1&2.989&3.02(13)&2.86(37)& &2.68(22)& &2.92(29)&3.03(24)&2.78(59)&  \\
\hline
6/1&3.465&3.58(17)&3.41(43)& & & &3.61(40)&3.37(25)&3.02(64)&  \\
\hline
7/1&3.940&4.07(18)&3.89(46)& & & &4.00(41)&4.02(28)&3.61(79)&  \\
\hline
8/1&4.415&4.56(21)&4.44(53)& & & &4.49(44)&4.47(30)& &  \\
\hline
\end{tabular}
\caption{Average eigenvalue ratios in the $\nu = 4$ sector in RMT and the $A_i$ and $B_i$ lattices.}\label{bigtable4}
\end{center}
\end{table}

\newpage

\begin{table}
\small
\begin{center}
\begin{tabular}{|ccccccccccc|}
\hline
$j/k$ & RMT & $B_1(S)$ & $B_2(S)$ & $B_1(F)$ & $B_2(F)$ & $A_1(S)$ & $A_2(S)$ & $A_3(S)$ & $A_4(S)$ & $A_5(S)$ \\
\hline
\hline
2/1&1.465& 1.53(8)& 1.45(9)&1.42(24)&1.41(24)& 1.45(9)&1.35(10)& 1.47(7)&1.46(13)&1.45(12) \\
\hline
3/1&1.897&1.99(12)&1.99(12)&1.85(31)&1.81(29)&1.97(13)&1.73(11)& 1.88(8)&1.88(15)&1.91(14) \\
\hline
4/1&2.316&2.55(16)&2.45(14)&2.21(35)&2.14(31)&2.33(14)&2.17(14)&2.27(10)&2.30(18)&2.30(17) \\
\hline
5/1&2.729&3.02(17)&2.94(17)& &2.25(44)& &2.55(17)&2.84(16)&2.71(21)&2.68(20) \\
\hline
6/1&3.139&3.47(21)&3.43(20)& & & &2.93(18)&3.27(18)&3.12(23)&3.08(22) \\
\hline
7/1&3.547&3.92(23)&3.90(23)& & & &3.29(20)&3.66(20)& &3.51(25) \\
\hline
8/1&3.952&4.47(26)&4.46(26)& & & &3.73(22)&4.07(21)& &3.84(26) \\
\hline
\end{tabular}
\caption{Average eigenvalue ratios in the $\nu = 5$ sector in RMT and the $A_i$ and $B_i$ lattices.}\label{bigtable5}
\end{center}
\end{table}

\newpage

\begin{table}
\small
\begin{center}
\begin{tabular}{|ccccccccccc|}
\hline
$j/k$ & RMT & $B_1(S)$ & $B_2(S)$ & $B_1(F)$ & $B_2(F)$ & $A_1(S)$ & $A_2(S)$ & $A_3(S)$ & $A_4(S)$ & $A_5(S)$ \\
\hline
\hline
2/1&1.417&1.44(10)&1.50(14)& & &1.44(12)&1.40(18)&1.47(21)& &  \\
\hline
3/1&1.801&1.83(13)&1.89(16)& & &1.82(14)&1.89(21)&1.84(24)& &  \\
\hline
4/1&2.172&2.26(16)&2.25(19)& & & &2.22(27)&2.36(31)& &  \\
\hline
5/1&2.537&2.67(17)&2.67(23)& & & &2.70(30)&2.62(38)& &  \\
\hline
6/1&2.898&3.08(20)&3.17(25)& & & &3.06(31)&3.05(52)& &  \\
\hline
7/1&3.256&3.59(23)&3.64(30)& & & &3.45(34)&3.28(48)& &  \\
\hline
8/1&3.612&4.00(25)&4.13(35)& & & &3.90(35)&3.72(48)& &  \\
\hline
\end{tabular}
\caption{Average eigenvalue ratios in the $\nu = 6$ sector in RMT and the $A_i$ and $B_i$ lattices.}\label{bigtable6}
\end{center}
\end{table}

\newpage

\begin{table}
\small
\begin{center}
\begin{tabular}{|ccccccccccc|}
\hline
$j/k$ & RMT & $B_1(S)$ & $B_2(S)$ & $B_1(F)$ & $B_2(F)$ & $A_1(S)$ & $A_2(S)$ & $A_3(S)$ & $A_4(S)$ & $A_5(S)$ \\
\hline
\hline
2/1&1.384&1.47(16)&1.47(14)& & & 1.38(7)& 1.39(9)&1.44(13)&1.39(20)&1.35(73) \\
\hline
3/1&1.732&1.76(17)&1.83(17)& & & 1.79(9)&1.78(10)&1.80(15)&1.78(21)&1.70(87) \\
\hline
4/1&2.067&2.17(22)&2.20(20)& & &2.53(14)&2.12(15)&2.00(17)&2.18(25)&1.9(1.0) \\
\hline
5/1&2.395&2.57(25)&2.62(25)& & &2.87(18)&2.45(16)&2.40(21)&2.47(29)&2.1(1.1) \\
\hline
6/1&2.719&3.00(27)&3.08(27)& & &3.34(21)&2.77(22)&2.77(24)& &2.6(1.3) \\
\hline
7/1&3.041&3.41(30)&3.50(30)& & &3.67(26)&3.12(25)&3.12(27)& &2.9(1.4) \\
\hline
8/1&3.360&3.79(33)&3.90(34)& & &3.91(29)&3.44(26)&3.51(28)& &3.2(1.4) \\
\hline
\end{tabular}
\caption{Average eigenvalue ratios in the $\nu = 7$ sector in RMT and the $A_i$ and $B_i$ lattices.}\label{bigtable7}
\end{center}
\end{table}

\newpage

\begin{table}
\small
\begin{center}
\begin{tabular}{|ccccccccccc|}
\hline
$j/k$ & RMT & $B_1(S)$ & $B_2(S)$ & $B_1(F)$ & $B_2(F)$ & $A_1(S)$ & $A_2(S)$ & $A_3(S)$ & $A_4(S)$ & $A_5(S)$ \\
\hline
\hline
2/1&1.353& 1.40(9)&1.28(17)& & &1.39(16)& 1.32(6)&1.37(13)&1.30(25)&  \\
\hline
3/1&1.671&1.83(10)&1.60(20)& & & & 1.63(7)&1.73(31)&1.69(32)&  \\
\hline
4/1&1.976&2.11(12)&2.02(25)& & & & 1.94(8)&2.16(33)&2.14(38)&  \\
\hline
5/1&2.274&2.50(14)&2.40(30)& & & &2.28(10)&2.39(33)&2.49(44)&  \\
\hline
6/1&2.567&2.92(16)&2.69(34)& & & &2.59(12)&2.70(40)& &  \\
\hline
7/1&2.859&3.24(17)&3.16(43)& & & &2.87(14)&3.05(47)& &  \\
\hline
8/1&3.148&3.73(22)&3.73(50)& & & & & & &  \\
\hline
\end{tabular}
\caption{Average eigenvalue ratios in the $\nu = 8$ sector in RMT and the $A_i$ and $B_i$ lattices.}\label{bigtable8}
\end{center}
\end{table}

\newpage

\begin{table}
\small
\begin{center}
\begin{tabular}{|ccccccccccc|}
\hline
$j/k$ & RMT & $B_1(S)$ & $B_2(S)$ & $B_1(F)$ & $B_2(F)$ & $A_1(S)$ & $A_2(S)$ & $A_3(S)$ & $A_4(S)$ & $A_5(S)$ \\
\hline
\hline
2/1&1.328&1.34(10)&1.20(13)& & & 1.35(9)&1.38(13)&1.35(14)& &  \\
\hline
3/1&1.623&1.68(12)&1.63(18)& & & &1.70(17)&1.55(16)& &  \\
\hline
4/1&1.904&1.97(13)&1.87(20)& & & &2.09(22)&1.79(19)& &  \\
\hline
5/1&2.178&2.28(14)&2.26(26)& & & &1.80(27)&2.08(20)& &  \\
\hline
6/1&2.448&2.59(17)&2.59(26)& & & &2.05(18)&2.38(22)& &  \\
\hline
7/1&2.714&2.98(19)&2.88(30)& & & &2.42(19)&2.68(25)& &  \\
\hline
8/1&2.979&3.48(30)&3.18(35)& & & & & & &  \\
\hline
\end{tabular}
\caption{Average eigenvalue ratios in the $\nu = 9$ sector in RMT and the $A_i$ and $B_i$ lattices.}\label{bigtable9}
\end{center}
\end{table}

\newpage

\begin{table}
\small
\begin{center}
\begin{tabular}{|ccccccccccc|}
\hline
$j/k$ & RMT & $B_1(S)$ & $B_2(S)$ & $B_1(F)$ & $B_2(F)$ & $A_1(S)$ & $A_2(S)$ & $A_3(S)$ & $A_4(S)$ & $A_5(S)$ \\
\hline
\hline
2/1&1.309&1.35(10)&1.30(11)& & & & 1.39(8)& 1.26(9)&1.36(11)&1.37(12) \\
\hline
3/1&1.583&1.74(13)&1.57(11)& & & & 1.71(9)& 1.52(9)&1.62(13)&1.64(12) \\
\hline
4/1&1.845&2.09(15)&1.86(13)& & & &1.96(10)&1.72(10)&1.90(14)&1.88(14) \\
\hline
5/1&2.099&2.40(17)&2.15(15)& & & &2.31(12)&2.08(11)& &2.10(16) \\
\hline
6/1&2.348&2.68(19)&2.49(16)& & & &2.59(13)&2.31(12)& &2.41(18) \\
\hline
7/1&2.594&3.00(22)&2.67(20)& & & & & & &2.63(19) \\
\hline
8/1&2.838&3.30(24)&2.97(22)& & & & & & &2.84(20) \\
\hline
\end{tabular}
\caption{Average eigenvalue ratios in the $\nu = 10$ sector in RMT and the $A_i$ and $B_i$ lattices.}\label{bigtable10}
\end{center}
\end{table}

\newpage

\begin{table}
\small
\begin{center}
\begin{tabular}{|ccccccccccc|}
\hline
$j/k$ & RMT & $B_1(S)$ & $B_2(S)$ & $B_1(F)$ & $B_2(F)$ & $A_1(S)$ & $A_2(S)$ & $A_3(S)$ & $A_4(S)$ & $A_5(S)$ \\
\hline
\hline
2/1&1.292&1.31(12)&1.36(18)& & & &1.50(23)& & &  \\
\hline
3/1&1.549&1.57(14)&1.76(22)& & & &1.72(26)& & &  \\
\hline
4/1&1.793&1.83(16)&2.05(26)& & & &2.03(29)& & &  \\
\hline
5/1&2.029&2.12(18)&2.31(27)& & & &2.32(33)& & &  \\
\hline
6/1&2.262&2.42(18)&2.62(31)& & & &2.64(38)& & &  \\
\hline
7/1&2.491&2.74(29)&2.80(32)& & & & & & &  \\
\hline
8/1&2.719&2.93(28)&3.11(35)& & & & & & &  \\
\hline
\end{tabular}
\caption{Average eigenvalue ratios in the $\nu = 11$ sector in RMT and the $A_i$ and $B_i$ lattices.}\label{bigtable11}
\end{center}
\end{table}

\newpage

\begin{table}
\small
\begin{center}
\begin{tabular}{|ccccccccccc|}
\hline
$j/k$ & RMT & $B_1(S)$ & $B_2(S)$ & $B_1(F)$ & $B_2(F)$ & $A_1(S)$ & $A_2(S)$ & $A_3(S)$ & $A_4(S)$ & $A_5(S)$ \\
\hline
\hline
2/1&1.275&1.34(14)&1.30(10)& & &1.29(11)&1.30(11)&1.29(19)& 1.47(8)&1.25(10) \\
\hline
3/1&1.518&1.62(16)&1.60(11)& & &1.54(12)&1.58(13)&1.50(21)&1.62(10)&1.43(14) \\
\hline
4/1&1.747&1.88(18)&1.90(13)& & &1.55(11)&1.79(14)&1.73(25)& &1.67(19) \\
\hline
5/1&1.969&2.14(20)&2.22(16)& & &1.76(12)&2.02(16)&1.94(24)& &1.83(24) \\
\hline
6/1&2.187&2.68(27)&2.46(19)& & & & & & &2.06(19) \\
\hline
7/1&2.401&2.92(29)&2.81(21)& & & & & & &2.26(17) \\
\hline
8/1&2.614&3.19(31)&3.08(24)& & & & & & &2.37(17) \\
\hline
\end{tabular}
\caption{Average eigenvalue ratios in the $\nu = 12$ sector in RMT and the $A_i$ and $B_i$ lattices.}\label{bigtable12}
\end{center}
\end{table}

\newpage

\begin{table}
\small
\begin{center}
\begin{tabular}{|ccccccccccc|}
\hline
$j/k$ & RMT & $B_1(S)$ & $B_2(S)$ & $B_1(F)$ & $B_2(F)$ & $A_1(S)$ & $A_2(S)$ & $A_3(S)$ & $A_4(S)$ & $A_5(S)$ \\
\hline
\hline
2/1&1.262&1.20(15)&1.29(14)& & &1.31(18)&1.32(13)&1.30(14)&1.16(29)&  \\
\hline
3/1&1.492&1.46(17)&1.55(16)& & &1.31(16)&1.65(16)&1.64(19)& &  \\
\hline
4/1&1.708&1.69(20)&1.83(21)& & &1.58(16)&1.87(18)&1.84(21)& &  \\
\hline
5/1&1.918&1.88(22)&2.13(23)& & &1.84(21)&2.12(19)&2.07(23)& &  \\
\hline
6/1&2.123&2.57(31)&2.35(24)& & & & & & &  \\
\hline
7/1&2.324&2.77(37)&2.57(26)& & & & & & &  \\
\hline
8/1&2.524&3.04(48)&2.80(28)& & & & & & &  \\
\hline
\end{tabular}
\caption{Average eigenvalue ratios in the $\nu = 13$ sector in RMT and the $A_i$ and $B_i$ lattices.}\label{bigtable13}
\end{center}
\end{table}

\newpage

\begin{table}
\small
\begin{center}
\begin{tabular}{|ccccccccccc|}
\hline
$j/k$ & RMT & $B_1(S)$ & $B_2(S)$ & $B_1(F)$ & $B_2(F)$ & $A_1(S)$ & $A_2(S)$ & $A_3(S)$ & $A_4(S)$ & $A_5(S)$ \\
\hline
\hline
2/1&1.250&1.24(14)&1.30(19)& & &1.36(19)&1.33(11)&1.16(41)& &  \\
\hline
3/1&1.468&1.57(17)&1.58(27)& & &1.79(22)&1.55(13)& & &  \\
\hline
4/1&1.674&1.88(20)&1.73(44)& & &2.13(29)&1.88(14)& & &  \\
\hline
5/1&1.872&2.18(38)& & & & & & & &  \\
\hline
6/1&2.067&2.48(42)& & & & & & & &  \\
\hline
7/1&2.257&2.68(46)& & & & & & & &  \\
\hline
8/1&2.446&2.91(49)& & & & & & & &  \\
\hline
9/1&2.633&3.17(57)& & & & & & & &  \\
\hline
3/2&1.175&1.27(14)&1.22(28)& & &1.32(11)&1.17(10)& & &  \\
\hline
4/2&1.339&1.52(17)&1.33(39)& & &1.56(16)&1.42(10)& & &  \\
\hline
5/2&1.498&1.76(31)& & & & & & & &  \\
\hline
6/2&1.653&2.00(34)& & & & & & & &  \\
\hline
7/2&1.806&2.16(38)& & & & & & & &  \\
\hline
8/2&1.956&2.35(40)& & & & & & & &  \\
\hline
9/2&2.106&2.56(47)& & & & & & & &  \\
\hline
4/3&1.140&1.20(12)&1.09(33)& & &1.19(10)& 1.21(8)& & &  \\
\hline
5/3&1.275&1.38(24)& & & & & & & &  \\
\hline
6/3&1.407&1.58(27)& & & & & & & &  \\
\hline
7/3&1.537&1.71(29)& & & & & & & &  \\
\hline
8/3&1.666&1.85(31)& & & & & & & &  \\
\hline
9/3&1.793&2.02(36)& & & & & & & &  \\
\hline
5/4&1.119&1.16(20)& & & & & & & &  \\
\hline
6/4&1.235&1.32(22)& & & & & & & &  \\
\hline
7/4&1.349&1.43(24)& & & & & & & &  \\
\hline
8/4&1.461&1.55(26)& & & & & & & &  \\
\hline
9/4&1.573&1.69(30)& & & & & & & &  \\
\hline
6/5&1.104&1.14(25)& & & & & & & &  \\
\hline
7/5&1.206&1.23(27)& & & & & & & &  \\
\hline
8/5&1.306&1.34(29)& & & & & & & &  \\
\hline
9/5&1.406&1.46(33)& & & & & & & &  \\
\hline
7/6&1.092&1.08(23)& & & & & & & &  \\
\hline
8/6&1.183&1.17(25)& & & & & & & &  \\
\hline
9/6&1.274&1.28(28)& & & & & & & &  \\
\hline
8/7&1.083&1.09(23)& & & & & & & &  \\
\hline
9/7&1.166&1.18(27)& & & & & & & &  \\
\hline
9/8&1.077&1.09(24)& & & & & & & &  \\
\hline
\end{tabular}
\caption{Average eigenvalue ratios in the $\nu = 14$ sector in RMT and the $A_i$ and $B_i$ lattices.}\label{bigtable14}
\end{center}
\end{table}

\newpage

\begin{table}
\small
\begin{center}
\begin{tabular}{|ccccccccccc|}
\hline
$j/k$ & RMT & $B_1(S)$ & $B_2(S)$ & $B_1(F)$ & $B_2(F)$ & $A_1(S)$ & $A_2(S)$ & $A_3(S)$ & $A_4(S)$ & $A_5(S)$ \\
\hline
\hline
2/1&1.240&1.32(10)&1.31(19)& & &1.19(11)&1.28(15)&1.48(16)&0.93(28)&1.16(23) \\
\hline
3/1&1.448&1.62(11)&1.55(21)& & &1.37(21)&1.48(16)&1.73(22)&1.12(26)&1.38(29) \\
\hline
4/1&1.644&1.88(12)&1.84(26)& & &1.66(20)&1.68(17)&1.94(23)&1.31(29)&1.50(28) \\
\hline
5/1&1.832&2.11(15)&2.11(30)& & & & & &1.55(30)&1.60(31) \\
\hline
6/1&2.016&2.36(16)&2.30(32)& & & & & &1.82(31)&1.72(33) \\
\hline
7/1&2.197&2.56(17)&2.56(35)& & & & & &2.09(44)&1.88(34) \\
\hline
8/1&2.376&2.78(17)&2.72(37)& & & & & &2.29(55)&2.00(35) \\
\hline
\end{tabular}
\caption{Average eigenvalue ratios in the $\nu = 15$ sector in RMT and the $A_i$ and $B_i$ lattices.}\label{bigtable15}
\end{center}
\end{table}

\newpage
\clearpage
\thispagestyle{empty}

\newpage
\clearpage

\newpage
\clearpage
\thispagestyle{empty}

\newpage
\clearpage

\newpage
\clearpage

\newpage
\clearpage

\newpage
\clearpage

\newpage
\clearpage

\newpage
\clearpage

\begin{thebibliography}{99}

\footnotesize

%%%%%%%%%%% weinberg, susskind %%%%%%%%%%%%%%%%%

%\cite{Weinberg:1975gm}
\bibitem{Weinberg:1975gm}
S.~Weinberg,
%``Implications Of Dynamical Symmetry Breaking,''
Phys.\ Rev.\  D {\bf 13} (1976) 974.
%%CITATION = PHRVA,D13,974;%%

%\cite{Susskind:1978ms}
\bibitem{Susskind:1978ms}
L.~Susskind,
%``Dynamics Of Spontaneous Symmetry Breaking In The Weinberg-Salam Theory,''
Phys.\ Rev.\  D {\bf 20}, 2619 (1979).
%%CITATION = PHRVA,D20,2619;%%

%\cite{Hill:2002ap}
\bibitem{Hill:2002ap}
C.~T.~Hill and E.~H.~Simmons,
%``Strong dynamics and electroweak symmetry breaking,''
Phys.\ Rept.\  {\bf 381}, 235 (2003)
[Erratum-ibid.\  {\bf 390}, 553 (2004)]
[arXiv:hep-ph/0203079].
%%CITATION = PRPLC,381,235;%%

%\cite{Sannino:2004qp}
\bibitem{Sannino:2004qp}
F.~Sannino and K.~Tuominen,
%``Techniorientifold,''
Phys.\ Rev.\  D {\bf 71}, 051901 (2005)
[arXiv:hep-ph/0405209].
%%CITATION = PHRVA,D71,051901;%%

%\cite{Dietrich:2006cm}
\bibitem{Dietrich:2006cm}
D.~D.~Dietrich and F.~Sannino,
%``Walking in the SU(N),''
Phys.\ Rev.\  D {\bf 75}, 085018 (2007)
[arXiv:hep-ph/0611341].
%%CITATION = PHRVA,D75,085018;%%

%%%%%%%%% neuberger %%%%%%%%%%%%%%%%%%%%%%%%%

%\cite{Neuberger:1997fp}
\bibitem{Neuberger:1997fp}
H.~Neuberger,
%``Exactly massless quarks on the lattice,''
Phys.\ Lett.\  B {\bf 417} (1998) 141
[arXiv:hep-lat/9707022].
%%CITATION = PHLTA,B417,141;%%

%%%%%%%%%%%%%%%%%%%%%%%%%%%%%%%%%%%%%%%%%%%%%%%%%%%%%

%\cite{Banks:1979yr}
\bibitem{Banks:1979yr}
T.~Banks and A.~Casher,
%``Chiral Symmetry Breaking In Confining Theories,''
Nucl.\ Phys.\  B {\bf 169}, 103 (1980).
%%CITATION = NUPHA,B169,103;%%


%%%%%%%%%%%%%%%%%% RMT %%%%%%%%%%%%%%%%%%%%%%%%%%%%

%\cite{Shuryak:1992pi}
\bibitem{Shuryak:1992pi}
E.~V.~Shuryak and J.~J.~M.~Verbaarschot,
%``Random Matrix Theory And Spectral Sum Rules For The Dirac Operator In
%QCD,''
Nucl.\ Phys.\  A {\bf 560}, 306 (1993)
[arXiv:hep-th/9212088].
%%CITATION = NUPHA,A560,306;%%

%\cite{Verbaarschot:1993pm}
\bibitem{Verbaarschot:1993pm}
J.~J.~M.~Verbaarschot and I.~Zahed,
%``Spectral density of the QCD Dirac operator near zero virtuality,''
Phys.\ Rev.\ Lett.\  {\bf 70}, 3852 (1993)
[arXiv:hep-th/9303012].
%%CITATION = PRLTA,70,3852;%%

%\cite{Verbaarschot:1994qf}
\bibitem{Verbaarschot:1994qf}
J.~J.~M.~Verbaarschot,
%``The Spectrum of the QCD Dirac operator and chiral random matrix theory: The
%Threefold way,''
Phys.\ Rev.\ Lett.\  {\bf 72}, 2531 (1994)
[arXiv:hep-th/9401059].
%%CITATION = PRLTA,72,2531;%%

%\cite{Halasz:1995qb}
\bibitem{Halasz:1995qb}
A.~M.~Halasz and J.~J.~M.~Verbaarschot,
%``Effective Lagrangians and chiral random matrix theory,''
Phys.\ Rev.\  D {\bf 52}, 2563 (1995)
[arXiv:hep-th/9502096].
%%CITATION = PHRVA,D52,2563;%%

%\cite{Fodor:2009nh}
\bibitem{Fodor:2009nh}
Z.~Fodor, K.~Holland, J.~Kuti, D.~Nogradi and C.~Schroeder,
%``Topology and higher dimensional representations,''
arXiv:0905.3586 [hep-lat].
%%CITATION = ARXIV:0905.3586;%%

%\cite{Damgaard:2001fg}
\bibitem{Damgaard:2001fg}
P.~H.~Damgaard, U.~M.~Heller, R.~Niclasen and B.~Svetitsky,
%``Patterns of spontaneous chiral symmetry breaking in vectorlike gauge
%theories,''
Nucl.\ Phys.\  B {\bf 633}, 97 (2002)
[arXiv:hep-lat/0110028].
%%CITATION = NUPHA,B633,97;%%

%\cite{Edwards:1999ra}
\bibitem{Edwards:1999ra}
R.~G.~Edwards, U.~M.~Heller, J.~E.~Kiskis and R.~Narayanan,
%``Quark spectra, topology and random matrix theory,''
Phys.\ Rev.\ Lett.\  {\bf 82}, 4188 (1999)
[arXiv:hep-th/9902117].
%%CITATION = PRLTA,82,4188;%%


%%%%%%%% old stuff %%%%%%%%%%%%%%%%%%%%%%%%%%%%%%%%%%%%%%%

%\cite{Kogut:1982fn}
\bibitem{Kogut:1982fn}
J.~B.~Kogut, M.~Stone, H.~W.~Wyld, J.~Shigemitsu, S.~H.~Shenker and D.~K.~Sinclair,
%``The Scales Of Chiral Symmetry Breaking In Quantum Chromodynamics,''
Phys.\ Rev.\ Lett.\  {\bf 48}, 1140 (1982).
%%CITATION = PRLTA,48,1140;%%

%\cite{Kogut:1984sb}
\bibitem{Kogut:1984sb}
J.~B.~Kogut, J.~Shigemitsu and D.~K.~Sinclair,
%``Chiral Symmetry Breaking With Octet And Sextet Quarks,''
Phys.\ Lett.\  B {\bf 145}, 239 (1984).
%%CITATION = PHLTA,B145,239;%%

%\cite{Iwasaki:1991mr}
\bibitem{Iwasaki:1991mr}
Y.~Iwasaki, K.~Kanaya, S.~Sakai and T.~Yoshie,
%``Quark confinement and number of flavors in strong coupling lattice QCD,''
Phys.\ Rev.\ Lett.\  {\bf 69}, 21 (1992).
%%CITATION = PRLTA,69,21;%%

%\cite{Damgaard:1997ut}
\bibitem{Damgaard:1997ut}
P.~H.~Damgaard, U.~M.~Heller, A.~Krasnitz and P.~Olesen,
%``On lattice QCD with many flavors,''
Phys.\ Lett.\  B {\bf 400}, 169 (1997)
[arXiv:hep-lat/9701008].
%%CITATION = PHLTA,B400,169;%%

%\cite{Karsch:1998qj}
\bibitem{Karsch:1998qj}
F.~Karsch and M.~Lutgemeier,
%``Deconfinement and chiral symmetry restoration in an SU(3) gauge theory
%with adjoint fermions,''
Nucl.\ Phys.\  B {\bf 550}, 449 (1999)
[arXiv:hep-lat/9812023].
%%CITATION = NUPHA,B550,449;%%


%\cite{Iwasaki:2003de}
\bibitem{Iwasaki:2003de}
Y.~Iwasaki, K.~Kanaya, S.~Kaya, S.~Sakai and T.~Yoshie,
%``Phase structure of lattice QCD for general number of flavors,''
Phys.\ Rev.\  D {\bf 69}, 014507 (2004)
[arXiv:hep-lat/0309159].
%%CITATION = PHRVA,D69,014507;%%



%%%%%%%% Sannino %%%%%%%%%%%%%%%%%%%%%%%%%%%%%%%%%%%%%%%%%

%%CITATION = PHRVA,D76,105004;%%

%\cite{Catterall:2007yx}
\bibitem{Catterall:2007yx}
S.~Catterall and F.~Sannino,
%``Minimal walking on
%the lattice,''
Phys.\ Rev.\  D {\bf 76}, 034504 (2007)
[arXiv:0705.1664 [hep-lat]].
%%CITATION =
%PHRVA,D76,034504;%%

%\cite{Appelquist:2007hu}
\bibitem{Appelquist:2007hu}
T.~Appelquist, G.~T.~Fleming and E.~T.~Neil,
%``Lattice Study of the Conformal Window in QCD-like Theories,''
Phys.\ Rev.\ Lett.\  {\bf 100}, 171607 (2008)
[arXiv:0712.0609 [hep-ph]].
%%CITATION = PRLTA,100,171607;%%


%%%%%%% Tom DeGrand and the Israelis %%%%%%%%%%%%%%%%%%%%%%%%

%\cite{DeGrand:2008kx}
\bibitem{DeGrand:2008kx}
T.~DeGrand, Y.~Shamir and B.~Svetitsky,
%``Phase structure of SU(3) gauge theory with two flavors of
%symmetric-representation fermions,''
Phys.\ Rev.\  D {\bf 79}, 034501 (2009)
[arXiv:0812.1427 [hep-lat]].
%%CITATION = PHRVA,D79,034501;%%

%\cite{Shamir:2008pb}
\bibitem{Shamir:2008pb}
Y.~Shamir, B.~Svetitsky and T.~DeGrand,
%``Zero of the discrete beta function in SU(3) lattice gauge theory with color
%sextet fermions,''
Phys.\ Rev.\  D {\bf 78}, 031502 (2008)
[arXiv:0803.1707 [hep-lat]].
%%CITATION = PHRVA,D78,031502;%%




%%%%%%%% Del Debbio gang %%%%%%%%%%%%%%%%%%%%%%%%%%%%%%%%%

%\cite{DelDebbio:2008tv}
\bibitem{DelDebbio:2008tv}
L.~Del Debbio, A.~Patella and C.~Pica,
%``Fermions in higher representations. Some results about SU(2) with adjoint
%fermions,''
arXiv:0812.0570 [hep-lat].
%%CITATION = ARXIV:0812.0570;%%

%\cite{DelDebbio:2008zf}
\bibitem{DelDebbio:2008zf}
L.~Del Debbio, A.~Patella and C.~Pica,
%``Higher representations on the lattice: numerical simulations. SU(2) with
%adjoint fermions,''
arXiv:0805.2058 [hep-lat].
%%CITATION = ARXIV:0805.2058;%%

%\cite{DelDebbio:2009fd}
\bibitem{DelDebbio:2009fd}
L.~Del Debbio, B.~Lucini, A.~Patella, C.~Pica and A.~Rago,
%``Conformal vs confining scenario in SU(2) with adjoint fermions,''
arXiv:0907.3896 [hep-lat].
%%CITATION = ARXIV:0907.3896;%%


%%%%%%%% us %%%%%%%%%%%%%%%%%%%%%%%%%%%%%%%%%%%%%%%%%%%%%%

%\cite{Fodor:2008hn}
\bibitem{Fodor:2008hn}
Z.~Fodor, K.~Holland, J.~Kuti, D.~Nogradi and C.~Schroeder,
%``Probing technicolor theories with staggered fermions,''
arXiv:0809.4890 [hep-lat].
%%CITATION = ARXIV:0809.4890;%%

%\cite{Fodor:2008hm}
\bibitem{Fodor:2008hm}
Z.~Fodor, K.~Holland, J.~Kuti, D.~Nogradi and C.~Schroeder,
%``Nearly conformal electroweak sector with chiral fermions,''
arXiv:0809.4888 [hep-lat].
%%CITATION = ARXIV:0809.4888;%%


%%%%%%%% Catteral, Giedt, Vranas %%%%%%%%%%%%%%%%%%%%%%%%%%%%%%%%%

%\cite{Catterall:2008qk}
\bibitem{Catterall:2008qk}
S.~Catterall, J.~Giedt, F.~Sannino and J.~Schneible,
%``Phase diagram of SU(2) with 2 flavors of dynamical adjoint quarks,''
JHEP {\bf 0811}, 009 (2008)
[arXiv:0807.0792 [hep-lat]].
%%CITATION = JHEPA,0811,009;%%

%\cite{Hietanen:2008mr}
\bibitem{Hietanen:2008mr}
A.~J.~Hietanen, J.~Rantaharju, K.~Rummukainen and K.~Tuominen,
%``Spectrum of SU(2) lattice gauge theory with two adjoint Dirac flavours,''
JHEP {\bf 0905}, 025 (2009)
[arXiv:0812.1467 [hep-lat]].
%%CITATION = JHEPA,0905,025;%%


%%%%%%%% Pallante gang %%%%%%%%%%%%%%%%%%%%%%%%%%%%%%%%%%%

%\cite{Deuzeman:2008sc}
\bibitem{Deuzeman:2008sc}
A.~Deuzeman, M.~P.~Lombardo and E.~Pallante,
%``The physics of eight flavours,''
Phys.\ Lett.\  B {\bf 670}, 41 (2008)
[arXiv:0804.2905 [hep-lat]].
%%CITATION = PHLTA,B670,41;%%

%\cite{Armoni:2008nq}
\bibitem{Armoni:2008nq}
A.~Armoni, B.~Lucini, A.~Patella and C.~Pica,
%``Lattice Study of Planar Equivalence: The Quark Condensate,''
Phys.\ Rev.\  D {\bf 78}, 045019 (2008)
[arXiv:0804.4501 [hep-th]].
%%CITATION = PHRVA,D78,045019;%%

%\cite{Jin:2008rc}
\bibitem{Jin:2008rc}
X.~Y.~Jin and R.~D.~Mawhinney,
%``Lattice QCD with Eight Degenerate Quark Flavors,''
PoS {\bf LATTICE2008}, 059 (2008)
[arXiv:0812.0413 [hep-lat]].
%%CITATION = POSCI,LATTICE2008,059;%%


%%%%%%%% Appelquist gang %%%%%%%%%%%%%%%%%%%%%%%%%%%%%%%%%

%\cite{Appelquist:2009ty}
\bibitem{Appelquist:2009ty}
T.~Appelquist, G.~T.~Fleming and E.~T.~Neil,
%``Lattice Study of Conformal Behavior in SU(3) Yang-Mills Theories,''
arXiv:0901.3766 [hep-ph].
%%CITATION = ARXIV:0901.3766;%%

%\cite{Hietanen:2009az}
\bibitem{Hietanen:2009az}
A.~J.~Hietanen, K.~Rummukainen and K.~Tuominen,
%``Evolution of the coupling constant in SU(2) lattice gauge theory with two
%adjoint fermions,''
arXiv:0904.0864 [hep-lat].
%%CITATION = ARXIV:0904.0864;%%


%%%%%%%% Finnish gang %%%%%%%%%%%%%%%%%%%%%%%%%%%%%%%%%%%%

%\cite{Hietanen:2009zz}
\bibitem{Hietanen:2009zz}
A.~Hietanen, J.~Rantaharju, K.~Rummukainen and K.~Tuominen,
%``Minimal technicolor on the lattice,''
Nucl.\ Phys.\  A {\bf 820} (2009) 191C.
%%CITATION = NUPHA,A820,191C;%%

%\cite{Deuzeman:2009mh}
\bibitem{Deuzeman:2009mh}
A.~Deuzeman, M.~P.~Lombardo and E.~Pallante,
%``Evidence for a conformal phase in SU(N) gauge theories,''
arXiv:0904.4662 [hep-ph].
%%CITATION = ARXIV:0904.4662;%%

%\cite{DeGrand:2009et}
\bibitem{DeGrand:2009et}
T.~DeGrand,
%``Volume scaling of Dirac eigenvalues in SU(3) lattice gauge theory with
%color sextet fermions,''
arXiv:0906.4543 [hep-lat].
%%CITATION = ARXIV:0906.4543;%%

%\cite{Hasenfratz:2009ea}
\bibitem{Hasenfratz:2009ea}
A.~Hasenfratz,
%``Investigating the critical properties of beyond-QCD theories using Monte
%Carlo Renormalization Group matching,''
arXiv:0907.0919 [hep-lat].
%%CITATION = ARXIV:0907.0919;%%

%\cite{Fodor:2009wk}
\bibitem{Fodor:2009wk}
Z.~Fodor, K.~Holland, J.~Kuti, D.~Nogradi and C.~Schroeder,
%``Nearly conformal gauge theories in finite volume,''
arXiv:0907.4562 [hep-lat].
%%CITATION = ARXIV:0907.4562;%%

%\cite{Nagai:2009ip}
\bibitem{Nagai:2009ip}
K.~i.~Nagai, G.~Carrillo-Ruiz, G.~Koleva and R.~Lewis,
%``Exploration of SU(N_c) gauge theory with many Wilson fermions at strong
%coupling,''
arXiv:0908.0166 [hep-lat].
%%CITATION = ARXIV:0908.0166;%%

%\cite{Dimopoulos:1979sp}
\bibitem{Dimopoulos:1979sp}
S.~Dimopoulos,
%``Technicolored Signatures,''
Nucl.\ Phys.\  B {\bf 168}, 69 (1980).
%%CITATION = NUPHA,B168,69;%%

%\cite{Osborn:1998qb}
\bibitem{Osborn:1998qb}
J.~C.~Osborn, D.~Toublan and J.~J.~M.~Verbaarschot,
%``From chiral random matrix theory to chiral perturbation theory,''
Nucl.\ Phys.\  B {\bf 540}, 317 (1999)
[arXiv:hep-th/9806110].
%%CITATION = NUPHA,B540,317;%%

%\cite{Nagao:2000qn}
\bibitem{Nagao:2000qn}
T.~Nagao and S.~M.~Nishigaki,
%``Massive chiral random matrix ensembles at beta = 1 and 4: Finite-volume
%QCD partition functions,''
Phys.\ Rev.\  D {\bf 62}, 065006 (2000)
[arXiv:hep-th/0001137].
%%CITATION = PHRVA,D62,065006;%%

%\cite{Damgaard:2000ah}
\bibitem{Damgaard:2000ah}
P.~H.~Damgaard and S.~M.~Nishigaki,
%``Distribution of the k-th smallest Dirac operator eigenvalue,''
Phys.\ Rev.\  D {\bf 63}, 045012 (2001)
[arXiv:hep-th/0006111].
%%CITATION = PHRVA,D63,045012;%%

%\cite{Damgaard:2000qt}
\bibitem{Damgaard:2000qt}
P.~H.~Damgaard, U.~M.~Heller, R.~Niclasen and K.~Rummukainen,
%``Eigenvalue distributions of the QCD Dirac operator,''
Phys.\ Lett.\  B {\bf 495}, 263 (2000)
[arXiv:hep-lat/0007041].
%%CITATION = PHLTA,B495,263;%%

%\cite{Akemann:1996vr}
\bibitem{Akemann:1996vr}
G.~Akemann, P.~H.~Damgaard, U.~Magnea and S.~Nishigaki,
%``Universality of random matrices in the microscopic limit and the Dirac
%operator spectrum,''
Nucl.\ Phys.\  B {\bf 487}, 721 (1997)
[arXiv:hep-th/9609174].
%%CITATION = NUPHA,B487,721;%%

%\cite{Damgaard:1997ye}
\bibitem{Damgaard:1997ye}
P.~H.~Damgaard and S.~M.~Nishigaki,
%``Universal spectral correlators and massive Dirac operators,''
Nucl.\ Phys.\  B {\bf 518}, 495 (1998)
[arXiv:hep-th/9711023].
%%CITATION = NUPHA,B518,495;%%

%\cite{Banks:1981nn}
\bibitem{Banks:1981nn}
T.~Banks and A.~Zaks,
%``On The Phase Structure Of Vector-Like Gauge Theories With Massless
%Fermions,''
Nucl.\ Phys.\  B {\bf 196}, 189 (1982).
%%CITATION = NUPHA,B196,189;%%

%\cite{Kovacs:2008sc}
\bibitem{Kovacs:2008sc}
T.~G.~Kovacs,
%``Gapless Dirac Spectrum at High Temperature,''
PoS {\bf LATTICE2008}, 198 (2008)
[arXiv:0810.4763 [hep-lat]].
%%CITATION = POSCI,LATTICE2008,198;%%

%\cite{Kovacs:2009zj}
\bibitem{Kovacs:2009zj}
T.~G.~Kovacs,
%``Absence of correlations in the QCD Dirac spectrum at high temperature,''
arXiv:0906.5373 [hep-lat].
%%CITATION = ARXIV:0906.5373;%%

%%%%%%% fundamental overlap RMT %%%%%%%%%%%%%%%%

%\cite{Giusti:2003gf}
\bibitem{Giusti:2003gf}
L.~Giusti, M.~Luscher, P.~Weisz and H.~Wittig,
%``Lattice QCD in the epsilon-regime and random matrix theory,''
JHEP {\bf 0311}, 023 (2003)
[arXiv:hep-lat/0309189].
%%CITATION = JHEPA,0311,023;%%

%\cite{Fukaya:2007yv}
\bibitem{Fukaya:2007yv}
H.~Fukaya {\it et al.},
%``Two-flavor lattice QCD in the epsilon-regime and chiral Random Matrix
%Theory,''
Phys.\ Rev.\  D {\bf 76}, 054503 (2007)
[arXiv:0705.3322 [hep-lat]].
%%CITATION = PHRVA,D76,054503;%%


%%%%%%%% quenched scale %%%%%%%%%%%%%%%

%\cite{Necco:2001xg}
\bibitem{Necco:2001xg}
S.~Necco and R.~Sommer,
%``The N(f) = 0 heavy quark potential from short to intermediate  distances,''
Nucl.\ Phys.\  B {\bf 622}, 328 (2002)
[arXiv:hep-lat/0108008].
%%CITATION = NUPHA,B622,328;%%

%%%%%%% stout %%%%%%%%%%%%%%%%%%

%\cite{Morningstar:2003gk}
\bibitem{Morningstar:2003gk}
C.~Morningstar and M.~J.~Peardon,
%``Analytic smearing of SU(3) link variables in lattice QCD,''
Phys.\ Rev.\  D {\bf 69}, 054501 (2004)
[arXiv:hep-lat/0311018].
%%CITATION = PHRVA,D69,054501;%%

%\cite{Fodor:2003bh}
\bibitem{Fodor:2003bh}
Z.~Fodor, S.~D.~Katz and K.~K.~Szabo,
%``Dynamical overlap fermions, results with hybrid Monte-Carlo algorithm,''
JHEP {\bf 0408}, 003 (2004)
[arXiv:hep-lat/0311010].
%%CITATION = JHEPA,0408,003;%%

%\cite{Egri:2006zm}
\bibitem{Egri:2006zm}
G.~I.~Egri, Z.~Fodor, C.~Hoelbling, S.~D.~Katz, D.~Nogradi and K.~K.~Szabo,
%``Lattice QCD as a video game,''
Comput.\ Phys.\ Commun.\  {\bf 177}, 631 (2007)
[arXiv:hep-lat/0611022].
%%CITATION = CPHCB,177,631;%%


\end{thebibliography}
\end{document}